\documentclass[lettersize,l]{IEEEtran}
\usepackage{amsfonts}
\usepackage{amsmath}
\usepackage{amssymb}
\usepackage{tabularx}
\usepackage{tikz}
\usepackage{array}
\usepackage{textcomp}
\usepackage{stfloats}
\usepackage{url}
\usepackage{verbatim}
\usepackage{graphicx}
\usepackage{subcaption}
\usepackage{orcidlink}
\usepackage{cleveref}
\usepackage{diagbox}
\usepackage{colortbl}
\usepackage{xcolor}
\def\BibTeX{{\rm B\kern-.05em{\sc i\kern-.025em b}\kern-.08em
    T\kern-.1667em\lower.7ex\hbox{E}\kern-.125emX}}
\usepackage{balance}
\usepackage{algorithm}
\usepackage{algorithmic}
\usepackage{cite}
\usepackage{makecell}
\usepackage{esint}
\usepackage{tabularray}

\usepackage{soul}
\usepackage{esvect}
\usepackage{mathtools}
\usepackage{physics}
\usepackage{gensymb}
\usepackage{tablefootnote}

\usepackage{bm}

\usepackage{soul}
\newcommand{\rev}[1]{\textcolor{black}{#1}}
\newcommand{\revtwo}[1]{\textcolor{black}{#1}}

\usetikzlibrary{calc}

\DeclareMathOperator{\sinc}{sinc}

\newcommand{\mbf}[1]{\mathbf{#1}}

\newcommand\doverline[1]{%
\tikz[baseline=(nodeAnchor.base)]{
    \node[inner sep=0] (nodeAnchor) {$#1$}; 
    \draw[line width=0.1ex,line cap=round] 
        ($(nodeAnchor.north west)+(0.0em,0.2ex)$) 
            --
        ($(nodeAnchor.north east)+(0.0em,0.2ex)$) 
        ($(nodeAnchor.north west)+(0.0em,0.5ex)$) 
            --
        ($(nodeAnchor.north east)+(0.0em,0.5ex)$) 
    ;
}}
\newcommand{\sigmarx}{\mbf{\Sigma}_\text{rx}}
\newcommand{\Js}{\vec{\mbf{J}}_s}
\newcommand{\Ms}{\vec{\mbf{M}}_s}
\newcommand{\nhat}{\hat{\mbf{n}}}
\newcommand{\Zse}{Z_{se}}
\newcommand{\Ysm}{Y_{sm}}
\newcommand{\Kem}{K_{em}}

\newcommand{\Efield}{\vec{\mbf{E}}}



\newcommand{\ie}{, i.e., }
\newcommand{\eg}{, e.g., }
\newcommand{\zrf}{\mbf{z}^\text{RF}}
\newcommand{\zant}{\mbf{z}^\text{ant}}
\newcommand{\Ghms}{\mbf{G}^\text{HMS}}
\newcommand{\Whms}{\mbf{W}^\text{HMS}}
\newcommand{\Phms}{\mbf{P}^\text{HMS}}
\newcommand{\ahms}{\mbf{t}^\text{HMS}}
\newcommand{\WD}{\mbf{W}^D}
\newcommand{\EE}{\text{EE}}
\newcommand{\pc}{P_\text{C}}

\crefformat{equation}{~(#2#1#3)}
\crefname{figure}{Fig.}{Figs.} 

\crefrangeformat{equation}{~(#3#1#4)-(#5#2#6)}

\begin{document}
\title{Energy Efficient Wireless Communications by Harnessing Huygens' Metasurfaces}
\author{Maryam Rezvani\orcidlink{https://orcid.org/0000-0002-4492-2547}, \textit{Graduate Student Member, IEEE}, Raviraj Adve\orcidlink{https://orcid.org/0000-0003-0224-2209}, \textit{Fellow, IEEE}
 , Akram bin Sediq\orcidlink{https://orcid.org/0000-0003-1260-2853}, \textit{Member, IEEE}, and Amr El-Keyi\orcidlink{https://orcid.org/0000-0003-2903-4055}, \textit{Member, IEEE}
\thanks{
Manuscript received August 2024; revised January 2025; revised March 2025; accepted April 2025. Date of publication XXX; date of current version April 2025. The associate editor coordinating the review of the manuscript and approving it for publication was XXXX. This work was supported by Ericsson Canada.\\ \indent 
Maryam Rezvani and Raviraj Adve are with the Department of Electrical and Computer Engineering, University of Toronto, Toronto, Ontario M5S 3G4, Canada (e-mail: mrezvani@ece.utoronto.ca; rsadve@comm.utoronto.ca). \\  
Akram bin Sediq and Amr El-Keyi are with Ericsson Canada, Kanata, ON K2K 2V6, Canada (e-mail: akram.bin.sediq@ericsson.com, amr.el-keyi@ericsson.com). \\ \indent 
This paper has supplementary downloadable material available at http://ieeexplore.ieee.org., provided by the author. The material includes supplemental material. Contact mrezvani@ece.utoronto.ca for further questions about this work.\\ \indent 
 Digital Object Identifier 
}}
\markboth{Journal of Selected Area in Information Theory,~Vol.~XX, No.~X, XXX~2025}%
{Energy Efficient Wireless Communications by Harnessing Huygens' Metasurfaces}

\maketitle
\begin{abstract}
	 Ambitions for the next generation of wireless communication include high data rates, low latency, ubiquitous access, ensuring sustainability (in terms of consumption of energy and natural resources), all while maintaining a reasonable level of implementation complexity. Achieving these goals necessitates reforms in cellular networks, specifically in the physical layer and antenna design. The deployment of transmissive metasurfaces at basestations (BSs) presents an appealing solution, enabling beamforming in the radiated wave domain, minimizing the need for energy-hungry RF chains. Among various metasurface-based antenna designs, we propose using Huygens’ metasurface-based antennas (HMAs) at BSs. Huygens’ metasurfaces offer an attractive solution for antennas because, by utilizing Huygens’ equivalence principle, they allow independent control over both the amplitude and phase of the transmitted electromagnetic wave. In this paper, we investigate the fundamental limits of HMAs in wireless networks by integrating electromagnetic theory and information theory within a unified analytical framework. Specifically, we model the unique electromagnetic characteristics of HMAs and incorporate them into an information-theoretic optimization framework to determine their maximum achievable sum rate. By formulating an optimization problem that captures the impact of HMA’s hardware constraints and electromagnetic properties, we quantify the channel capacity of HMA-assisted systems. We then compare the performance of HMAs against phased arrays and other metasurface-based antennas in both rich scattering and realistic 3GPP channels, highlighting their potential in improving spectral and energy efficiency.
\end{abstract}


\section{Introduction}
\IEEEPARstart{S}{ustainability}, energy efficiency, reduction in consumption of natural resources while also providing high data rates, low latency, and ubiquitous access are among the main goals for the next generation of wireless communications~\cite{6Gvision}. Meeting all these demands requires reevaluating all aspects of cellular networks, including, of special interest here, redesigning the antenna at the core  of th network.

One proposed approach to providing high throughputs has been to use extra-large multiple-input multiple-output (XL-MIMO) antenna arrays, also known as digital phased arrays (DPAs)~\cite{XLMIMO}. In this approach, thousands of antennas are deployed at the basestations (BSs) with each antenna connected to a radio frequency (RF) chain. While this excessive number of RF chains and degrees of freedom (DoFs) in the system offers high data rates and flexibility, it also results in high power consumption, i.e., low energy efficiency. Additionally, XL-MIMO requires significant use of natural resources due to the large number of RF chains and semiconductor products used in its structure. Considering these aspects, it becomes evident that XL-MIMO is not a suitable choice for the next generation of wireless communications.

To reduce the power consumption of XL-MIMO, hybrid MIMO arrays\ie hybrid phased arrays (HPAs), have been introduced; here, part of the processing is pushed from the digital domain to the analog domain. In HPAs, a large number of antennas are connected to a relatively smaller number of RF chains through analog phase-shifters (PSs)~\cite{HeathHYB}. While the reduced number of RF chains results in lower power consumption (compared with DPAs) the PSs add to the latency of the system~\cite{PSpassive}. Additionally, the PSs, as well as the employed power combiners and dividers have non-trvial insertion losses, which degrade performance~\cite{HYBinserionLossWilkinson}. The larger delay and signal losses of HPAs motivates us to explore alternative antenna designs.

Metasurface (MTS) antennas, where part of the signal processing is transformed to the \rev{radiated} electromagnetic (EM) wave domain, are a relatively recent proposal for use in wireless networks. Specifically, the use of MTSs in the form of reconfigurable intelligent surfaces (RISs) to control the wireless communication channel is a recently introduced paradigm~\cite{RISDiRenzo}. An RIS is a passive MTS composed of subwavelength unit-cells, where the properties of each unit-cell can be manipulated by applying a proper voltage to the implemented PIN diodes or varactors. By deploying reflective RISs between a BS and its users, the reflected EM wave is directed toward the user, enhancing communication quality. 

Transmissive RISs can also be deployed at the BS~\cite{eleftheriades2022prospects} providing the possibility of forming \textit{any} desired radiation pattern at the EM level, i.e., ``layer-0"~\cite{RISDiRenzo}. \rev{By properly configuring a transmissive RIS deployed in front of a small antenna array, we can achieve performance comparable to that of a large DPA, such as massive MIMO systems, while using a limited number of RF chains.} This substantial reduction, without the need for adding analog devices, will likely extend lifecycles, reduce the risk of failure, and hence decrease maintenance costs. Importantly, the production of MTSs uses the same technology as printed circuit boards~\cite{eleftheriades2022prospects}, lowering manufacturing costs. In sum, high energy efficiency, the reduced need for RF chains, and limited costs of MTS antennas make them a strong candidate for the next generation of wireless communications. In what follows, we focus on MTS antennas and provide a detailed description of their operating principle.

\rev{When an EM wave excites an MTS, the interaction between the wave and the unit-cells induces electric and magnetic surface currents, whose characteristics depend on the unit-cell configuration. These currents form a boundary condition and radiate a modified EM wave~\cite[Ch.~7.3]{balanis2012advanced}. Since MTSs are much thinner than the operating wavelength, this interaction effectively creates a discontinuity in the EM wave at the macroscopic level~\cite{eleftheriades2022prospects, selvanayagam2013discontinuous}.} There are two main types of MTS antennas that differ in their excitation technique: one is waveguide-fed, in which the MTS transforms a waveguide mode to free-space radiation~\cite{SmithTermination2018dynamically}\eg dynamic metasurface antennas (DMAs). Alternatively, one or more transmissive MTSs can be placed in front of a DPA and, by manipulating the properties of the propagating signal, form the desired radiation pattern\eg stacked intelligent MTS antennas (SIMAs)~\cite{SIM_ICC}.

A DMA is constructed by a few microstrip lines, spaced by $\lambda/2$ ($\lambda$ = wavelength), where each microstrip line is loaded with subwavelength unit-cells~\cite{Alex1_2021dynamic}. Each unit-cell is loosely coupled to the corresponding microstrip line, and the desired radiation pattern is formed by sampling from the EM wave in the waveguide and radiating into free space (downlink) or vice versa (uplink)~\cite{DMACE}. In a DMA, the weighting factor of each element either follows a Lorentzian phase modulation or binary amplitude modulation~\cite{DMA2017Analysis}. This link between the phase and amplitude modulation is a major limitation of DMAs as it limits the possible radiation patterns. \revtwo{Furthermore, due to the needed loose coupling between unit-cells and microstrip lines, in the uplink, a significant portion of the signal power does not transfer to the waveguide mode, resulting in reduced energy  efficiency~\cite{DMAuplinkEfficiency}}. 

In contrast, a SIMA comprises several cascaded MTS layers and manipulates the impinging signal by changing its phase while propagating through each layer. The SIMA structure has been introduced in~\cite{Jarahi2018allSIM} as a diffractive deep neural network which operates at the speed of light. Notably, however, the multilayer structure of the SIMA introduces losses, as each MTS layer has a power transmission coefficient less than one~\cite{SIMuplinkEfficiency}. Furthermore, although the SIMA does not need any digital combiner and forms the desired radiation pattern in the wave domain only, its complex structure requires complex signal processing techniques, usually in form of a gradient ascent/descent algorithm~\cite{SIM_ICC,SIM_JSAC,SIMuplinkEfficiency}. 

\begin{figure}[t]
	\centering
	\begin{subfigure}{\linewidth}
		\centering
		\includegraphics[width=\linewidth]{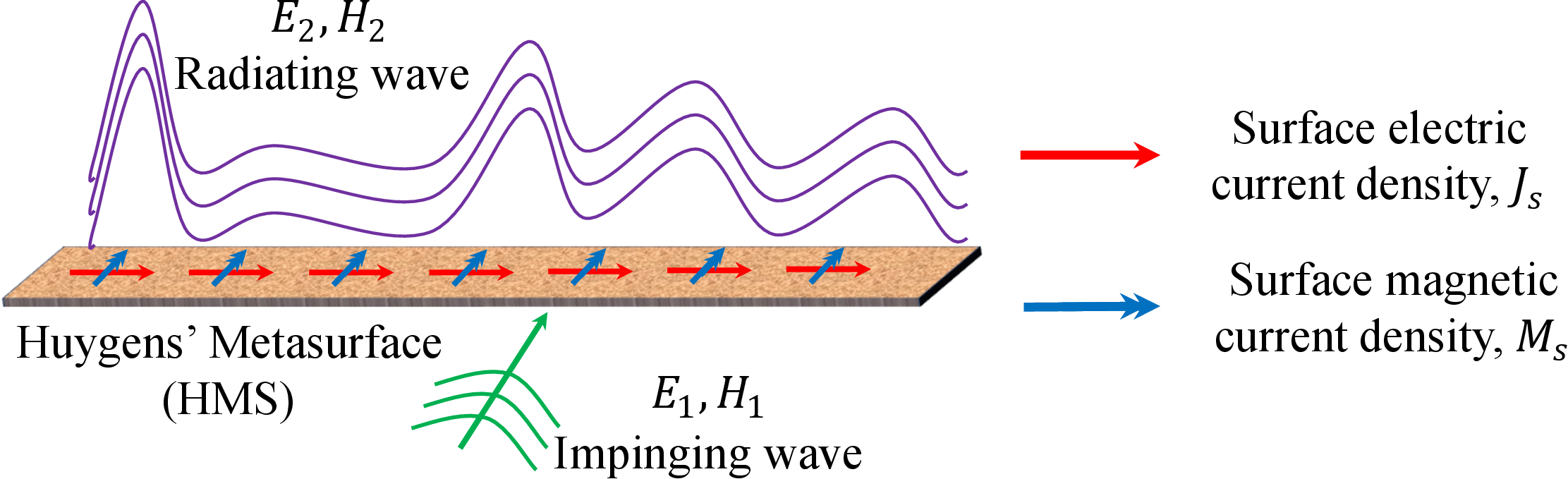} 
		\caption{}
		\vspace{2mm}
		\label{fig:Huygensprinciple}
	\end{subfigure}
	\hfill
	\begin{subfigure}{0.8\linewidth}
		\centering
		\includegraphics[width=\linewidth]{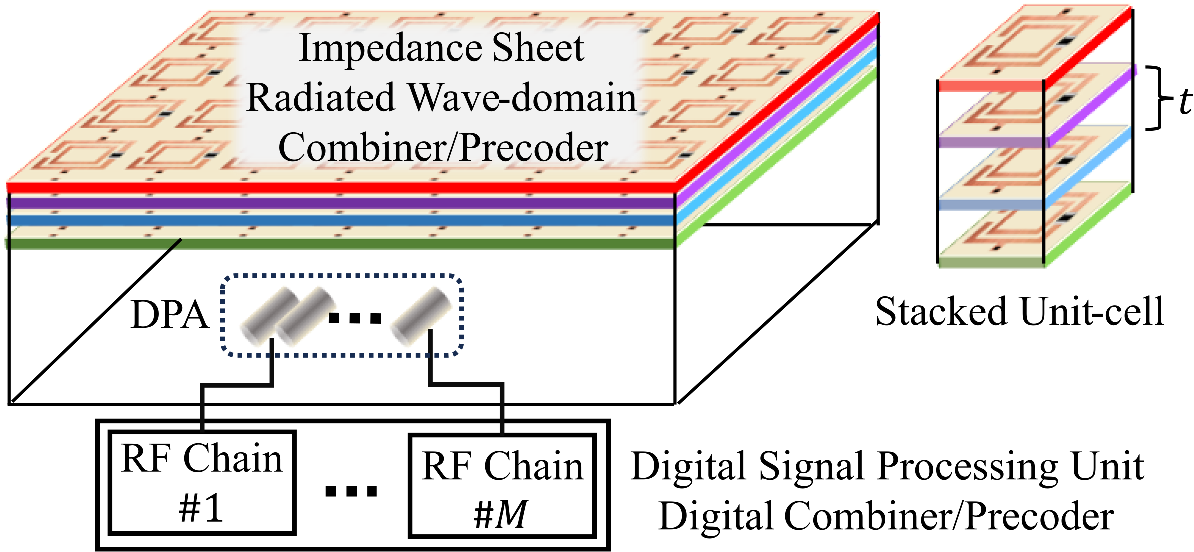} 
		\caption{}
		\label{fig:HMSunitcell}
	\end{subfigure}
	\caption{(a) Huygens' principle (b) structure of HMA comprising HMS and DPA and the detailed structure of a stacked unit-cell }
	\label{fig:HMS_all}
\end{figure}

The limited control over the DMA's weighting factor and complex structure of SIMA motivates us to propose deploying Huygens' metasurface (HMS) antennas at the BS. An HMS is a thin surface composed of subwavelength unit-cells that, in theory, can transform any impinging EM wave into any desired EM radiation. The working principle of an HMS is the surface equivalence theorem~\cite{Eleftheriades2018theoryHMS} introduced by Schelkunoff as a generalization of Huygens' principle. The surface equivalence theorem uses the EM uniqueness theorem stating that \textit{"the EM radiation in a lossy and closed region can be fully described by the tangential electric and magnetic fields over the boundary and electric and magnetic sources within that region"}~\cite[Ch.~7.8]{balanis2012advanced}. As illustrated in Fig.~\ref{fig:Huygensprinciple}, through the interaction between incident wave\ie EM fields in the first region, and the unit-cells of the HMS, needed electric and magnetic current densities are induced on the HMS to satisfy the desired discontinuity between the EM fields in the first and second regions. Therefore, the needed properties of HMS are dictated by the choice for the desired radiated and incident fields.

The HMS's unit-cells are made of co-located wire loops that mimic the behavior of Huygens' sources\footnote{Defined as orthogonal magnetic and electric dipoles~\cite{Eleftheriades2022HuygenSource}.}~\cite{Eleftheriades2022HuygenSource}. Hence, an HMS can independently control the magnitude and phase of the radiated field~\cite{Eleftheriades2021IndependentControl} and form any desired radiation pattern when used as an antenna. To achieve this, each unit-cell is formed by stacking four impedance sheets\footnote{Stacking three impedance sheets is enough to satisfy boundary conditions but stacking four offers full amplitude and phase control~\cite{Eleftheriades3vs4layers}.} separated by distance $t\ll \lambda$ (see Fig.~\ref{fig:HMSunitcell})~\cite{Eleftheriades2016HMSDesign}. Each unit-cell is controlled by applying an appropriate voltage to the implemented varactors. To form an antenna structure based on an HMS, a relatively large HMS layer is deployed in front of a DPA with a limited number of RF chains, as shown in Fig.~\ref{fig:HMSunitcell}. \rev{ In this setup, the HMS provides a radiated wave-domain (RWD) combiner (precoder) through the interaction between radiated waves from the users in the uplink (radiated wave from the DPA in the downlink) and unit-cells of the HMS. Digital signal processing is employed in the digital domain, where a digital combiner (precoder) processes the signals received (transmitted) by the DPA with its limited dimensionality and RF chains}. 

The independent control over amplitude and phase of the weighting factor of each unit-cell, its reduced size, and simple hardware makes HMS an ideal choice for deployment at BSs. \rev{Furthermore, its flexibility in forming any desired radiation can be used to enhance the cell coverage and provide adaptive beamforming~\cite{eleftheriades2022prospects, ataloglou2023metasurfaces}}. To show the superiority of HMS-based antennas (HMAs), we compare the achievable sum rate and energy efficiency (EE)\footnote{Defined as the ratio of the achievable sum rate to the power consumed.} of HMA with the popular antenna alternatives of DPA, HPA, DMA, and SIMA. To do so, using EM theory, we propose a model for an HMA that includes the hardware limitations of this type of antenna including various noise sources and insertion losses of the HMS. 

\rev{To effectively compare the EE of different antenna designs, it is essential to develop an accurate model that captures the power consumption of each system. Power consumption in HMA and other MTS-based antennas can be categorized into two main components: the RF chain and the HMS (or any type of transmissive MTS). The power consumption of RF chains has been extensively studied in the literature (e.g., see~\cite{RFPErkip, RFPchoi}), and the power consumption of reflective MTSs has been addressed in~\cite{Dai2023RISPwrConsump, wang2022RISPwrMeasu}. For MTS-based antennas, \cite{DMAEE} calculates the power consumption of a DMA, while \cite{Di2023PwrConsumpRRS} proposes a model for transmissive RIS-based antennas, accounting for the power consumed by the necessary controller circuitry. However, a comprehensive model that fully captures the power consumption of all components, including the RF chain and MTS, is still lacking. To fill this gap, we build upon the power consumption model presented in~\cite{DMAuplinkEfficiency}, incorporating a detailed breakdown of the RF chain and MTS driver circuitry. This model is based on simplified block diagrams and data from commercially available wireless communication devices, particularly those used in BSs.}

\rev{\textbf{Contributions and Motivations}: Driven by the goal of achieving sustainable, high-performance wireless communication, we propose Huygens' metasurface-based antennas (HMAs) as the antenna solution for the next generation of wireless systems. The primary objective of this paper is to determine the maximum achievable sum rate\ie optimum channel capacity, of HMA-assisted wireless communication systems in the uplink. To accomplish this, we integrate information theory (IT), which defines the limits of reliable information transfer~\cite{shannon1948mathematical}, with EM theory, which governs the behavior of the HMSs. This approach leads to the development of a novel system model and optimization framework that aligns with the principles of EM information theory (EIT). Our contributions in this paper include:}
\begin{itemize}
	\item We propose the use of Huygens' metasurface-based antennas for the next generation of wireless communications where high performance, energy efficiency, low latency, and cost efficiency are all required. 
	\item \rev{Leveraging both EM theory and information theory, we develop a detailed noise model, HMS model, and corresponding system model. This system model, along with the accompanying optimization problem to maximize channel capacity of HMA-assisted systems, accounts for the unique hardware and electromagnetic limitations of HMAs, by effectively combining IT and EM theory.} Also, by thoroughly studying the hardware requirements and circuitry of an MTS-based antenna, we propose a power consumption model that takes into account various sources of energy consumption. 
	\item We develop a fractional programming-based algorithm to optimize the achievable sum rate in HMA-assisted systems. Additionally, we propose a simple algorithm to find an effective combiner when serving a single user.
	\item We compare the uplink performance of an HMA with that of a DPA, HPA, DMA, and SIMA, using an equal aperture area, in terms of achievable sum rate and energy efficiency. We do so in rich scattering channels and realistic 3GPP channels generated by QuadRiGa~\cite{quadriga}.
\end{itemize}

Our results show that, with equal aperture areas, an HMA, while offering simple hardware implementation, achieves sum rates comparable to that of a DPA and SIMA, while surpassing the performance of DMAs. Moreover, HMA’s power consumption aligns with other MTS-based antennas, such as DMA and SIMA; as a result, HMAs have the highest energy efficiency among different antenna designs.

\textbf{Organization}: This paper is organized as follows: We begin by describing the structure of HMAs in Section~\ref{sec:HMA}; this section also presents its working principle and our proposed model for HMAs, accounting for its hardware limitations. We then discuss our system model, capturing the effect of various noise sources in the system, and formulate the optimization of achievable sum rate problem in HMA-assisted systems in Section~\ref{sec:sysmodelPF}. In Section~\ref{sec:WDFP}, we describe the use of fractional programming and zero-forcing to obtain the sum rate in a HMA-assisted systems. Then, we describe our power consumption model for calculating the energy efficiency of HMA in Section~\ref{sec:PWR}. Our numerical results are presented in Section~\ref{sec:numerical}. Finally, Section~\ref{sec:conclusion} summarizes and concludes the paper.

\textbf{Notation}: We represent vector fields and currents by $\vec{\mbf{A}}$; a hat ($\hat{\cdot}$) denotes a unit vector. We use lowercase and uppercase boldface letters to denote vectors and matrices, respectively. $(\cdot)^*$, $(\cdot)^H$, and $(\cdot)^{-1}$ denote conjugate, Hermitian and matrix inversion, respectively. Also, $\oslash$ and $\odot$ represents element-wise division and product. Furthermore, by $a_{i,j}$, we refer to the element on the $i$th row and $j$th column of the matrix $\mathbf{A}$ and, by $v_i$, the $i$th element of vector $\mathbf{v}$, respectively.  $\mathbf{A}_{k,:}$ denotes the $k$-th row of $\mathbf{A}$. Also, $\jmath$ is defined as $\sqrt{-1}$ and $\sinc(x) = \frac{\sin(\pi x)}{\pi x}$. Finally, $\mathcal{CN}(\boldsymbol{\mu}, \mathbf{R})$ represents the complex Gaussian distribution with mean $\boldsymbol{\mu}$ and covariance matrix $\mathbf{R}$, and by $\mbf{I}_N$, we refer to the $N \times N$ identity matrix.

\section{Huygens' Metasurface Antennas}~\label{sec:HMA}
In this section, we describe the working principle of HMSs from EM theory point of view, and then, propose a simplified model for HMAs which accounts for HMS's EM properties and captures the hardware limitations of HMAs. While our analysis of achievable rate is focused on the uplink, the simplest description of an HMA is as a radiating antenna\ie the downlink. Due to reciprocity, the HMA acting as a receiver operates similarly. Specifically, the reciprocity holds in this case because the HMS consists of passive and reciprocal unit-cells~\cite{reciprocity2016synthesis}, which we effectively model using a transfer coefficient. Additionally, the propagation medium between the DPA and the HMS is effectively free-space which is linear and isotropic. Therefore, the reciprocity principle holds in the HMA structure.

\subsection{Working Principle of Huygens' Metasurface}\label{sec:HMSWP}
\begin{figure}[t]
	\centering
	\includegraphics[width= 0.9\linewidth]{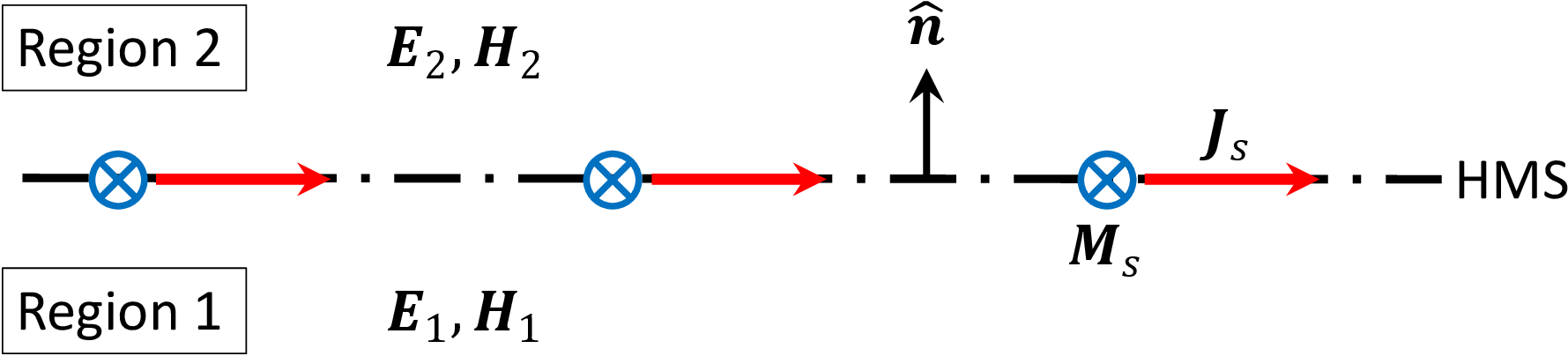}
	\caption{A macroscopic view of HMS, the induced $\Js$ and $\Ms$ supports the desired discontinuity in EM field.}
	\label{fig:BoundaryCond}
\end{figure}
At a high level, due to its extremely thin nature, the HMS is modeled by the appropriate boundary conditions needed to support the discontinuity of the EM field directly above and below the HMS. As depicted in Fig.~\ref{fig:BoundaryCond}, the HMS supports induced \textit{orthogonal} electric and magnetic current densities, $\Js$ and $\Ms$, respectively, leading to the boundary conditions~\cite{Vasilis2021HMSreview}
\begin{equation}
	\begin{split}
		\Js &= \nhat\times (\vec{\mbf{H}}_2 - \vec{\mbf{H}}_1)\\
		\Ms &= -\nhat\times (\vec{\mbf{E}}_2 - \vec{\mbf{E}}_1)
	\end{split}
	\label{eq:boundaryCond}
\end{equation}
where $\nhat$ is the unit normal vector at the boundary pointing toward region 2, and $\vec{\mbf{H}}_i$ and $\vec{\mbf{E}}_i$ are the magnetic and electric field vectors in the region $i$ , $i \in \{1,2\}$, respectively. Note that each pair of EM fields must satisfy Maxwell's equations~\cite{Vasilis2021HMSreview}. 

In a passive HMS, each unit-cell is designed such that the EM fields on both sides of the HMS induce $\Js$ and $\Ms$ to support the desired discontinuity in the EM field. To do so, another set of relations between the currents and the EM fields is introduced by defining an electric (magnetic) surface impedance (admittance) which relates the average tangential electric (magnetic) field on the HMS to the $\Js$ ($\Ms$)~\cite{Vasilis2021HMSreview}\ie
\begin{equation}
	\begin{split}
		\frac{1}{2} (\vec{\mbf{E}}_\text{t,1}+\vec{\mbf{E}}_\text{t,2}) &= \doverline{Z}_{se}\ .\ \Js -\doverline{K}_{em}\ .\ (\nhat \times \Ms)\\
		\frac{1}{2} (\vec{\mbf{H}}_\text{t,1}+\vec{\mbf{H}}_\text{t,2}) &= \doverline{Y}_{sm}\ .\ \Ms +\doverline{K}_{em}\ .\ (\nhat \times \Js)
	\end{split}
	\label{eq:tensorHMS}
\end{equation}
where $\doverline{Z}_{se}$ is the surface electric impedance tensor, $\doverline{Y}_{sm}$ is the surface magnetic admittance tensor, $\doverline{K}_{em}$ is the magnetoelectric coupling coefficient tensor, and $\vec{\mbf{E}}_{t,i}$ and $\vec{\mbf{H}}_{t,i}$ are the electric and magnetic fields tangential to the HMS surface $i\in\{1,2\}$. Note that by introducing $\doverline{K}_{em}$ in (\ref{eq:tensorHMS}), we consider the HMS to be bianisotropic, meaning that electric (magnetic) field on one side of the HMS can excite a magnetic (electric) field on the other side. This is a generalization of the non-bianisotropic HMS and by setting $\Kem$ to zero in (\ref{eq:tensorHMS}), we reach the non-bianisotropic HMS relations~\cite{Eleftheriades2018theoryHMS}.

By combining (\ref{eq:boundaryCond}) and (\ref{eq:tensorHMS}), and simplifying the relations by considering a transverse electric (TE) field\ie the electric field is orthogonal to the direction of propagation\footnote{This assumption does not affect our analysis since the wave receive by the HMA (in the uplink) is planar - the antenna is in the far-field of users and hence, the $\vec{\mbf{E}}$ and $\vec{\mbf{H}}$ fields are both transverse to the direction of propagation.}, we have
\begin{equation}
	\begin{split}
		\frac{1}{2} (\vec{\mbf{E}}_\text{t,1} + \vec{\mbf{E}}_\text{t,2}) = & \, \Zse (\nhat \times (\vec{\mbf{H}}_2 - \vec{\mbf{H}}_1)) \\
		& - \Kem (\nhat \times (-\nhat \times (\vec{\mbf{E}}_2 - \vec{\mbf{E}}_1))) \\
		\frac{1}{2} (\vec{\mbf{H}}_\text{t,1} + \vec{\mbf{H}}_\text{t,2}) = & \, \Ysm (-\nhat \times (\vec{\mbf{E}}_2 - \vec{\mbf{E}}_1)) \\
		& + \Kem (\nhat \times (\nhat \times (\vec{\mbf{H}}_2 - \vec{\mbf{H}}_1)))
	\end{split}
	\label{eq:impHMS}
\end{equation}

where, since the E-field is in one direction, $\Zse$, $\Ysm$, and $\Kem$ are scalars. All the variables in (\ref{eq:impHMS}) are a function of the location on the HMS\ie considering an HMS in the $z-x$ plane where $\nhat$ point in the $y$-direction, all variables would be a function of $x$ and $z$.

It has been shown that any field transformation can be achieved by a passive and lossless HMS, if the real parts of the power density orthogonal to the HMS surface on both sides are equal~\cite{Vasilis2021HMSreview}\ie
\begin{equation}
	\frac{1}{2} \text{Re}\{ \nhat\ . \ (\vec{\mbf{E}}_1\times \vec{\mbf{H}}_1^*)\} = \frac{1}{2} \text{Re}\{ \nhat\ . \ (\vec{\mbf{E}}_2 \times \vec{\mbf{H}}_2^*)\}.
	\label{eq:localPowerConservation}
\end{equation}
This condition is known as the local power conservation rule and must be satisfied at the center of each unit-cell. The local power conservation rule can be broken by introducing power distributing methods on the lower facet of HMS\eg using surface waves~\cite{vasilis2020nonlocalpower}, putting HMS on top of a cavity~\cite{HMSonCavity}, or using a pair of HMSs~\cite{pairofHMS}. Using these methods, we can transform the local power conservation rule to a global power conservation rule, such that the \textit{total} power normal to the surface of the HMS, in both regions, are equal\ie
\begin{equation}
	\small
	\frac{1}{2}\text{Re}\left\{\oiint_{HMS} \vec{\mbf{E}}_1 \times \vec{\mbf{H}}_1^* \, \cdot ds \right\} = 
	\frac{1}{2} \text{Re}\left\{\oiint_{HMS} \vec{\mbf{E}}_2 \times \vec{\mbf{H}}_2^* \, \cdot ds \right\}.
	\label{eq:GlobalPowerConservation}
\end{equation}

At a high level, under the global conservation rule, an HMS can introduce local gains and losses to the propagating EM wave. 
After choosing the desired pair of $\vec{\mbf{E}}_2$ and $\vec{\mbf{H}}_2$ which satisfies~\eqref{eq:GlobalPowerConservation} with respect to $\vec{\mbf{E}}_1$ and $\vec{\mbf{H}}_1$, the HMS properties\ie $\Zse$, $\Ysm$, and $\Kem$, are uniquely defined. This procedure is known as macroscopic level design~\cite{Eleftheriades2018theoryHMS}.

The next step is to discretize the macroscopic properties of the HMS over each unit-cell, leading to the microscopic level design. Each unit-cell is designed to have properties as dictated by the macroscopic level design~\cite{Eleftheriades2016HMSDesign} and hence, the required applied voltage to each varactor in each unit-cell is determined. This step can be completed by checking a pre-generated lookup table~\cite{HMSLookuptable} which relates the properties of each unit-cell\ie $\Kem$, $\Zse$, and $\Ysm$ to the required voltage bias for the varactor in the design\footnote{For a detailed discussion on steps of microscopic level design, the interested reader can refer to~\cite{Eleftheriades2018theoryHMS}.}.

In summary, if the global power conservation rule is satisfied by redistributing the incident EM power, an HMS can achieve any form of EM radiation needed to match the desired output power distribution. The redistributing process helps the HMS resemble local loss and gain behavior while the total power transmission coefficient of the HMS remains equal to one. Importantly, the HMS can offer independent control over amplitude and phase of the transmission coefficient of each unit-cell~\cite{Eleftheriades2021IndependentControl}. We note that the number of unit-cells and the proportional number of samples offered by an HMS is a key enabler in forming any desired radiation pattern by the HMS. The denser the unit-cells, the higher DoFs available in supporting the desired radiation pattern.

The above explanation on the working principle of an HMS prepares us to present our model for an HMA. Our model will, then, allow us to determine the performance of an HMA in a wireless network.

\subsection{Modeling Huygens' Metasurface Antennas}\label{sec:HMSmodel}
\begin{figure}[t]
	\centering
	\includegraphics[width = \linewidth]{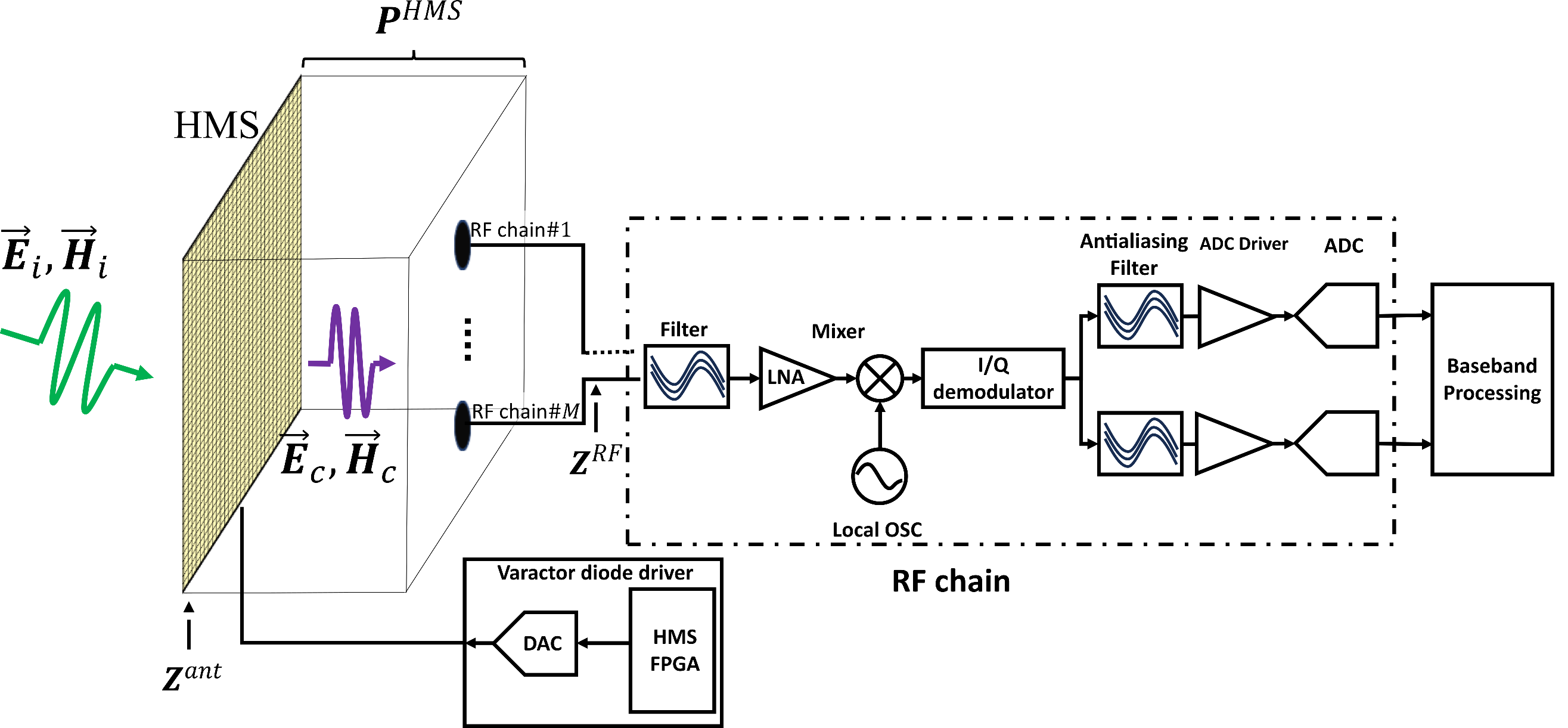}
	\caption{HMA structure: an HMS in front of a DPA with a limited number of antennas.}
	\label{fig:HMA}
\end{figure}

As illustrated in Fig.~\ref{fig:HMA}, an HMA comprises an HMS with $N$ unit-cells deployed in front of a DPA with $M$ RF chains where $M\ll N$. We assume that each element in the DPA is connected to an RF chain, though the DPA can be replaced by an HPA, partially or fully-connected, to further reduce the number of RF chains. Also, specific circuitry is needed to derive the required voltage to configure each unit-cell. 

In what follows, we model the effect of an HMA and its RWD combiner on the incident/source EM wave in the uplink using EM theory; the same procedure can be applied to downlink by the virtue of reciprocity theorem~\cite[Ch.~7.5]{balanis2012advanced}. To do so, we separate the RWD combiner into two parts, $\Phms$ and $\Whms$; the former captures the effect of EM propagation between HMS and the DPA, while the latter captures the effect of the weighting factor of each unit-cell in the HMS. We begin with $\Whms$.

In the uplink, assume that the incident EM wave on the $n$-th antenna\ie $n$-th unit-cell in HMS, is described as
\begin{equation}
	\vec{\mbf{E}}_{\text{t},\text{i},n}\Big|_{\vec{\mbf{r}}_n \in S_1} = \sqrt{P_T} s f(\vec{\mbf{r}}_n), \  \vec{\mbf{H}}_{\text{t},\text{i},n}\Big|_{\vec{\mbf{r}}_n \in S_1} = \frac{\sqrt{P_T}}{\eta_0} s f(\vec{\mbf{r}}_n)
\end{equation}
where $\vec{\mbf{E}}_{\text{t},\text{i},n}$ and $\vec{\mbf{H}}_{\text{t},\text{i},n}$ are the tangential components of the incident electric and magnetic fields on the $n$-th unit-cell, respectively, $P_T$ is the incident power, $s$ is the transmitted symbol, $\eta_0 $ is the free space wave impedance, $f(\vec{\mbf{r}})$ is a function which captures the dependence of the EM field on the location denoted by $\vec{\mbf{r}}$, and $S_1$ is the user-facing HMS surface. Upon incidence, the HMS couples $\vec{\mbf{E}}_\text{i}$ and $\vec{\mbf{H}}_\text{i}$ to the HMA structure and transforms it to a form determined by the excited electric and magnetic current densities on the HMS. We denote to this coupled EM wave as $\vec{\mbf{E}}_\text{c}$ and $\vec{\mbf{H}}_\text{c}$.

Since an HMS can support any form of discontinuity in the propagating EM wave, we can describe $\vec{\mbf{E}}_{\text{t},\text{c},n}$ and $\vec{\mbf{H}}_{\text{t},\text{c},n}$, the tangential components of $\vec{\mbf{E}}_\text{c}$ and $\vec{\mbf{H}}_\text{c}$ at the $n$-th unit cell, as
\begin{equation}
	\vec{\mbf{E}}_{\text{t},\text{c},n}\Big|_{\vec{\mbf{r}}_n \in S_2} = \sqrt{P_T} s g(\vec{\mbf{r}}_n), \  \vec{\mbf{H}}_{\text{t},\text{c},n}\Big|_{\vec{\mbf{r}}_n \in S_2} = \sqrt{P_T} s q(\vec{\mbf{r}}_n)
\end{equation}
where $S_2$ is the DPA-facing HMS surface. Note that $\vec{\mbf{E}}_{\text{c}}$ and $\vec{\mbf{H}}_{\text{c}}$ describe the desired propagation to the DPA while satisfying the Maxwell's equations~\cite{Vasilis2021HMSreview}; hence, $g(\vec{\mbf{r}})$ and $q(\vec{\mbf{r}})$ describe the tangential components of the EM field on $S_2$.

A simple choice for $g(\vec{\mbf{r}})$ and $q(\vec{\mbf{r}})$ is to be equal to a transformed version of $f(\vec{\mbf{r}}_n)$\ie $g(\vec{\mbf{r}}) = \eta_0q(\vec{\mbf{r}}) = t_n f(\vec{\mbf{r}}_n)$ where $t_n \in \mathbb{C}$ is the transmission coefficient equivalent to the forward path scattering parameter of the $n$-th unit-cell~\cite{Eleftheriades2021IndependentControl}. While choosing the coupled EM wave in this form restricts the HMA's performance, since we limit the available choices for $g(\vec{\mbf{r}})$ and $q(\vec{\mbf{r}})$ to being proportional to $f(\vec{\mbf{r}})$, it also simplifies the modeling process. This allows for a systematic analysis and makes sure the chosen $\vec{\mbf{E}}_{\text{c},n}$ and $\vec{\mbf{H}}_{\text{c},n}$ pair satisfies Maxwell's equations without delving into intricate design details\footnote{Since $t_n$ is constant over the $n$-th unit-cell, the following relations hold: $\nabla \cdot \vec{\mbf{E}}{\text{c},n} = t_n \nabla \cdot \vec{\mbf{E}}{\text{i},n}$, $\nabla \times \vec{\mbf{E}}{\text{c},n} = t_n \nabla \times \vec{\mbf{E}}{\text{i},n}$, $\nabla \cdot \vec{\mbf{H}}{\text{c},n} = t_n \nabla \cdot \vec{\mbf{H}}{\text{i},n}$, and $\nabla \times \vec{\mbf{H}}{\text{c},n} = t_n \nabla \times \vec{\mbf{H}}{\text{i},n}$. Therefore, Maxwell's equations are satisfied for the coupled electromagnetic wave since the incident wave satisfies Maxwell's equations over each unit-cell.}.

Under this assumption for the function form, we have
\begin{equation}
	\vec{\mbf{E}}_{c} = \ahms\odot \vec{\mbf{E}}_{i}, \ \ \ \vec{\mbf{H}}_{c} = \ahms\odot \vec{\mbf{H}}_{i}
\end{equation}
where $\ahms \in \mathbb{C}^{N \times 1}$ is the vector of transmission coefficients. To make sure the required $\vec{\mbf{E}}_{c}$ and $\vec{\mbf{H}}_{c}$ are producible by the HMS, the global power conservation rule in \eqref{eq:GlobalPowerConservation} must be satisfied. This constraint can be written as:
\begin{equation}
	\sum_{n=1}^{N}P_{i,n} =\sum_{n=1}^{N}P_{c,n}= \sum_{n=1}^{N}\abs{\ahms_n}^2 P_{i,n}
	\label{eq:HMSconstr}
\end{equation}
where $P_{x,n} = \frac{1}{2} \text{Re}\{\vec{\mbf{E}}_{x,n}\times (\vec{\mbf{H}}_{x,n})^*\} \forall x\in \{{i}, {c}\}$. Finally, circling back to the RWD combiner, the effect of weighting factor of unit-cells is modeled as 
\begin{equation}
	\Whms = \text{diag}(\ahms).
\end{equation}

We now model the effect of EM field propagation between the deployed DPA and the HMS. To do so, we apply the reciprocity theorem~\cite[Ch.~7.5]{balanis2012advanced}, effectively modeling EM propagation in the reverse direction. The specific form of the propagation matrix from the DPA to the HMS depends on the type of antenna elements in DPA and their corresponding Green’s function~\cite{HMSonCavity}. For illustrative purposes, here we assume that the DPA comprises Hertzian dipoles positioned in front of a perfect magnetic conductor\footnote{We emphasize that this is a choice and any phased array can be used.}. It is common practice to use this arrangement to enhance dipole radiation directivity and leverage the additional radiation gain due to image theory~\cite[Ch.~7.4]{balanis2012advanced}. We account for variations in the amplitude of the incident EM wave by capturing relative changes in EM power. For a Hertzian dipole, the Poynting vector can be described in both near-field and far-field regions as~\cite[Ch.~4.2]{balanis2016antenna}
%
\begin{equation}
	\begin{split}
		\vec{\mbf{W}}=& \frac{\eta_0 }{8} \abs{\frac{I_0 l_d}{\lambda_0 }}^2 \frac{\sin^2{\theta}}{r^2}\left[1-\jmath \frac{1}{(k_0 r)^3}\right]\hat{\mathbf{r}}\\
		&+\jmath \eta_0  \frac{k_0 |I_0l_d|^2\cos{\theta}\sin{\theta}}{16 \pi^2 r^3}[1+\frac{1}{(k_0 r)^2}]\hat{\boldsymbol{\theta}}\label{eq:dipoleW}
	\end{split}
\end{equation}
where $k_0 = \frac{2 \pi}{\lambda_0}$ is the free space wavenumber, $I_0$ is the excitation current, $l_d$ is the dipole length, $\eta_0 $ is the free space wave impedance, and $r$ and $\theta$ are the distance and azimuthal angle between source and evaluation point, respectively. Since, in this paper, our focus is on the uplink communication where BS is in the far field of the users, only the propagating term of the Poynting vector\ie its real part, is important.

Since the relative change in EM power due to propagation is important, we only account for the term depending on the EM field evaluation location\ie $\frac{\sin^2{\theta}}{r^2}$. Then, to ensure that the total power on a half-sphere at distance $r$ remains equal to the power radiated from the source\footnote{Since the propagation environment is equivalent to free space and the HMS within a sub-wavelength distance of the DPA, there are no propagation losses from each Hertzian dipole antenna to the HMS.}, we normalize $\frac{\sin^2{\theta}}{r^2}$ on this half sphere. Further, we map the power coefficient to the complex amplitude of $\vec{\mbf{E}}$, reaching the following relation for the propagation of the electric field from the $m$-th element of the DPA to the $n$-th unit-cell on the HMS:
\begin{equation}
	|\Phms_{m,n}| = \left(\frac{3\sqrt{2}}{8\pi}\frac{\sin^2{\theta_{m,n}}}{r^2_{m,n}}\right)^{\frac{1}{2}}.
	\label{eq:PHMSamp}
\end{equation}

To capture the change in phase of the electric field with respect to the source, we use the propagating term of the electric field radiated by a Hertzian dipole~\cite[Ch.~4.2]{balanis2016antenna}:
\begin{equation}          
	\vec{\mbf{E}} = \jmath \eta_0  I_0 l_d \frac{k_0 \sin{\theta}}{4\pi r}\left[1+\frac{1}{\jmath k_0 r}-\frac{1}{(k_0 r)^2}\right]e^{-\jmath k_0 r}\hat{\boldsymbol{\theta}}. \label{eq:dipoleE}
\end{equation}

The phase change experienced by the electric field due to the propagation from the $m$-th DPA element to the $n$-th unit-cell in the HMS is equal to the change of phase in $\vec{\mbf{E}}$ as described in~\eqref{eq:dipoleE} when $r = r_{m,n}$ (where $r_{m,n}$ is the distance between $m$-th antenna element and $n$-th unit-cell).  Hence, 
\begin{equation}
	\begin{split}
		\measuredangle \Phms_{m,n} =& \measuredangle \vec{\mbf{E}}_{m,n} + \frac{\pi}{2}- k_0 r_{m,n}\\
		&+\measuredangle\left(1+\frac{1}{\jmath k_0 r_{m,n}}-\frac{1}{(k_0 r_{m,n})^2}\right).
	\end{split}
	\label{eq:PHMSphase}
\end{equation}

Finally, combining~\eqref{eq:PHMSamp} and~\eqref{eq:PHMSphase} the $(m,n)$-th element of the propagation matrix $\Phms$ is given by $|\Phms_{m,n}|e^{\jmath \measuredangle \Phms_{m,n}}$. 

\textit{Insertion Loss of HMA}: Although, theoretically, the HMS is lossless, practical implementations incur losses due to non-idealities such as propagation in the dielectric substrate, copper losses in the unit cells, and the resonant nature of each unit-cell~\cite{Eleftheriades2018theoryHMS}. Beside HMS, additional power losses depend on the method used to satisfy the global power conservation rule in the HMA, as described in (5). For example, power losses are larger when the required power distribution is achieved through the excitation of surface waves~\cite{vasilis2020nonlocalpower}.
	
To account for the corresponding losses, we define an average insertion loss coefficient for HMS, $T^\text{HMS}$ and set $T^{\text{HMS}} = 0.7$, the worst-case value for our purpose among values reported in~\cite{ataloglou2023metasurfaces}. Note that multiple configurations of an HMS can potentially produce the same desired radiation pattern, but with varying power losses, as the non-idealities of each unit cell depend on its configuration. Therefore, in future work, the power losses of the HMS can be incorporated as an optimization constraint to determine the optimal configuration with the lowest losses~\cite{ataloglou2023metasurfaces}.

In sum, the complete RWD combiner in HMA is  
\begin{equation}
	\Ghms = \sqrt{T^\text{HMS}}\Phms\Whms.
\end{equation}

\section{System Model and Problem Formulation}\label{sec:sysmodelPF}
\rev{Building upon the developed model for the HMA, we now present our system model, which integrates principles of EM theory and information theory to analyze the performance of HMA-assisted communication systems. This model captures the effects of various noise sources unique to MTS-based systems and quantifies their impact on reliable information transfer. Using this model, we formulate an optimization problem to maximize the achievable sum rate in the uplink, a key metric that reflects the maximum channel capacity of the system.}

\begin{figure}
	\centering
	\includegraphics[width= 0.75\linewidth]{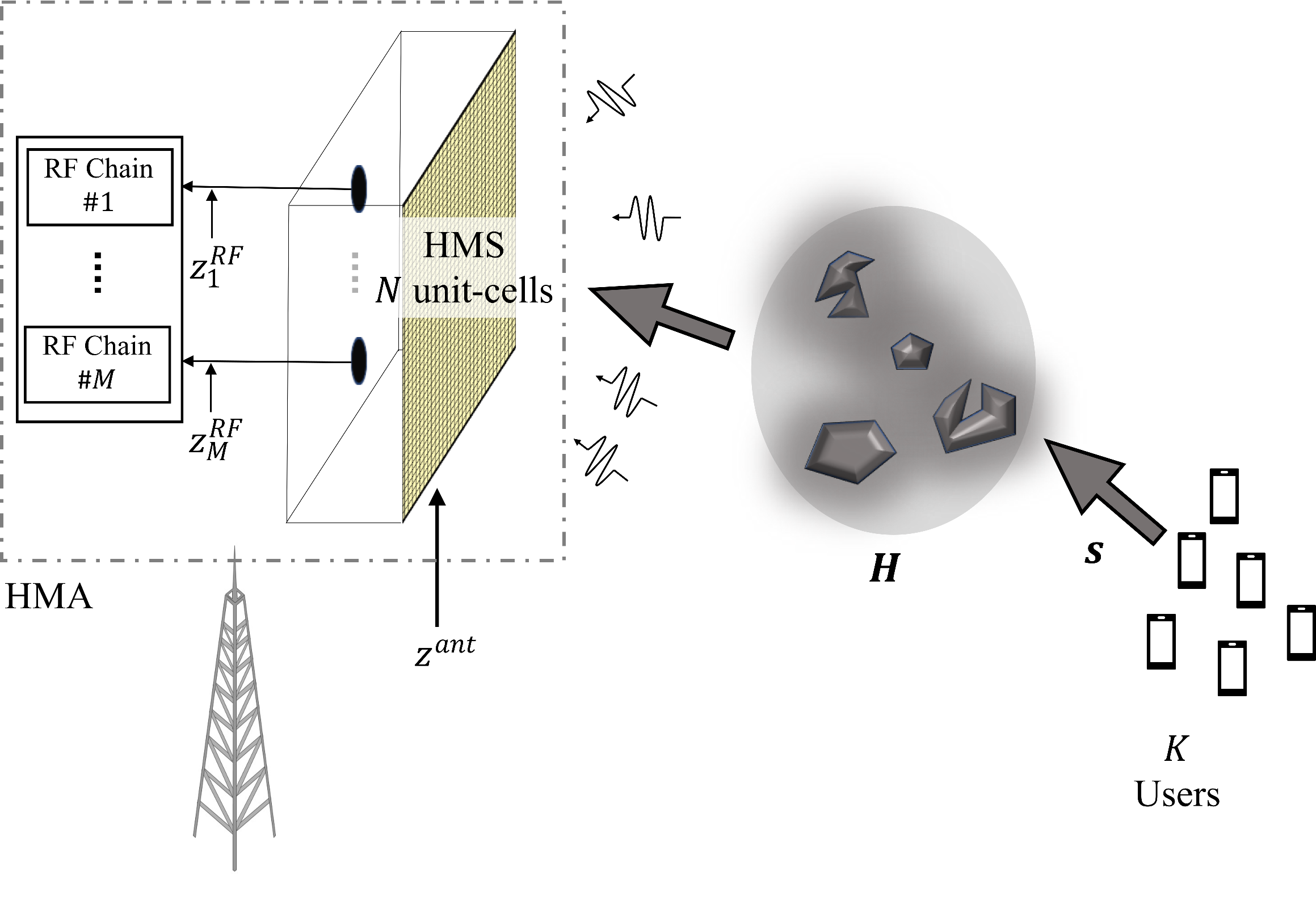}
	\caption{Uplink scenario}
	\label{fig:uplinksys}
\end{figure}

As shown in Fig.~\ref{fig:uplinksys}, in our model, a BS equipped with an HMA compromised of $N$ unit-cells and $M$ RF chains serves $K$ users. The signal $\mbf{x} \in \mathbb{C}^{N \times 1}$ incident on the \textit{antenna surface} is described by 
\begin{equation}
	\mbf{x} = \mbf{Hs}+\zant
\end{equation}
where $\mbf{H} \in \mathbb{C}^{N \times K}$ is the channel matrix, and $\zant$ is the external noise captured by the antenna where its entries are correlated by the same correlation matrix as the antenna~\cite{DMACE}.

The first stage of signal processing on the signal, $\mbf{x}$, happens at the EM level\ie layer-0, by passing thorough the HMS. Then the RF chain noise is added and the RF chains translate the signal to the digital domain and digital combining takes place. Therefore, the complete system model is described as:
\begin{equation}
	\mbf{y}=g\mbf{W}^D (\Ghms \mbf{x} + \zrf),
\end{equation}
where $\mbf{W}^D \in \mathbb{C}^{K \times M}$ is the digital combiner at baseband, $\Ghms$ is the RWD combiner introduced by the HMS, $g$ is the gain introduced by the RF chain, and $\zrf$ is the $M \times 1$ white noise vector added by the RF chains with power $\sigma_\text{RF}^2$\ie $\zrf \sim \mathcal{CN}(0,\sigma_\text{RF}^2\mbf{I}_M)$.

The optimum achievable sum rate in this system is defined as 
$R = \sum_{k=1}^K \log_2(1+\gamma_k)$
where $\gamma_k$ is the signal-to-interference-plus-noise ratio (SINR) for user $k$ defined as 
{\small
\begin{align}
	&\gamma_k= \notag \\
	\small
	&\frac{P_{T,k}\abs{\mbf{G}_{k,:}\mbf{H}_{:,k}}^2}{\sum_{j\neq k}^K P_{T,j}\abs{\mbf{G}_{k,:}\mbf{H}_{:,j}}^2 +\sigma_\text{ant}^2(\mbf{G}_{k,:}\mbf{\Sigma}_\text{rx}(\mbf{G}_{k,:})^H)+\sigma_\text{RF}^2|\mbf{W}^D_{k,:}|^2}.
	\label{eq:gammak}
\end{align}}%
where $P_{T,k}$ is the transmitted power from user $k$, $\sigma_\text{ant}^2$ is the external noise power captured by each unit-cell, $\mbf{G} =\mbf{W}^D \Ghms$ is the total combiner introduced by the HMA, and $\mbf{\Sigma}_\text{rx}$ is receiver's\ie antenna's, correlation matrix. We note two noise terms in denominator of~\eqref{eq:gammak}, where the first is the noise captured by the antenna from the environment colored by the RWD combiner and the second term is the commonly used white noise term added by the RF chain. Accounting for antenna noise in MTS-based antennas is crucial for a proper analysis~\cite{DMACE}. MTSs in these systems act as lenses, focusing the signal and noise captured from the environment onto a limited number of RF chains; this focusing results in a strong colored noise at the input of each RF chain.

To maximize $R$, we are interested in solving the following optimization problem:
\begin{subequations}
	\label{OP1:HMS}
	\begin{align}
		&\max_{\ahms, \mbf{W}^D} R\\
		\text{s.t.} \ \ \ &  \mbf{W}^\text{HMS} = \text{diag}(\ahms)\label{OP1cnstr:Wdef}\\
		&\sum_{n=1}^{N}P_{i,n} = \sum_{n=1}^{N}\abs{\ahms_n}^2 P_{i,n}
		\label{OP1cnstr:pwrconservation}      
	\end{align}
\end{subequations}
where~\eqref{OP1cnstr:Wdef} is the definition of the RWD combiner and~\eqref{OP1cnstr:pwrconservation} is the global power conservation rule of the HMS as defined in~\eqref{eq:HMSconstr}. In what follows, using EM theory, we rewrite~\eqref{OP1cnstr:pwrconservation} to capture the dependency of the signal power received by the HMS on the communication channel and transmitted power by the users.

The transmitted power by each user, $P_{T,k}$ is the amplitude of the Poynting vector of the EM field radiated by the corresponding user\ie $P_{T,k} =  \frac{\abs{\Efield_{\text{UE},k}}}{2\eta_0}\ [\frac{W}{m^2}]$. The user's transmitted signal undergoes the effect of the channel modeled by $\mbf{H}$ which changes the amplitude of both electric and magnetic fields. Therefore, the signal power captured by the $n$-th unit-cell of HMS from $k$-th user is equal to
\begin{equation}
	P_{\text{i},n,k} = \frac{P_{T,k}A_{\text{eff},n}}{2\eta_0} \abs{\mbf{h}_{n,k} \Efield_{\text{UE},k}}^2,\label{eq:Pin}
\end{equation}
where $A_{\text{eff},n}$ is the effective area of the $n$-th unit-cell of HMS. Since each unit-cell is approximated by a continuous current sheet, $A_{\text{eff},n}$ is equal to the physical area of each unit-cell~\cite[Ch.~2.15]{balanis2016antenna}. Therefore, the total power received by the $n$-th unit-cell from all $K$ users is equal to $P_{\text{i},n} = \sum_{k=1}^K P_{\text{i},n,k}$. Also, using~\eqref{eq:HMSconstr}, the total power received from all users after transmission thorough the HMS is equal to 
\begin{equation}
	P_{\text{c},n} =  \abs{\ahms_{n}}^2 P_{\text{i},n}.\label{eq:Pc}
\end{equation}

In this paper, we assume all the users transmit with equal power\ie $P_{T,k} = P_T$ and leave power scheduling as future work; also, all unit-cells have equal area\ie $A_{\text{eff},n} = A_\text{UC}$. Finally, by combining~\eqref{eq:Pin},~\eqref{eq:Pc}, and~\eqref{eq:HMSconstr}, we rewrite~\eqref{OP1cnstr:pwrconservation} as
\begin{equation}
	\norm{\text{vec}(\mbf{H})}^2 = \sum_{k=1}^K \norm{\ahms \odot \mbf{H}_{:,k}}^2.\label{eq:HMSfinalcnstr}
\end{equation}
The expression in~\eqref{eq:HMSfinalcnstr} elucidates the effect of communication channel on the received power by HMS and, in turn, on the power conservation rule of the HMS.

\section{Wave-Domain Combiner Design}\label{sec:WDFP}
In this section, our objective is to solve the optimization problem outlined in~\eqref{OP1:HMS}. To achieve this, we initially convert both the objective function and constraints into convex forms. Subsequently, we introduce a fractional programming-based algorithm to find a close-to-optimum RWD combiner. Additionally, when the HMA serves a single-antenna user, we develop a straightforward maximum ratio combiner.

The objective function in~\eqref{OP1:HMS} consists of a sum of non-decreasing functions of ratios, allowing a fractional programming (FP) reformulation. While FP is not inherently convex, we address this by employing the quadratic transform. As shown in~\cite{YUFP}, optimizing a sequence of non-decreasing functions of ratios $\frac{A_k(\mbf{x})}{B_k(\mbf{x})}$\ie 
\begin{subequations}
	\label{OP:FPorg}
	\begin{align}
		\max_\mbf{x} \sum_{k=1}^K &f_k\left( \frac{A_k(\mbf{x})}{B_k(\mbf{x})}\right)\\
		\text{s.t.} \ \ \ & \mbf{x} \in \mathcal{X},
	\end{align}
\end{subequations}
is equivalent to the following optimization problem:
\begin{subequations}
	\label{OP:Qtrans}
	\begin{align}
		\max_{\mbf{x}, \mbf{v}} \sum_{k=1}^K &f_k\left( 2v_k \sqrt{A_k(\mbf{x})}-v_k^2B_k(\mbf{x})\right)\\
		\text{s.t.} \ \ \ & \mbf{x} \in \mathcal{X}, \ v_k \in \mathbb{R}\ \ k=1, \dots,K,
	\end{align}
\end{subequations}
where the numerator and denominator in objective function of~\eqref{OP:FPorg} are decoupled using a quadratic transform. As shown in~\cite{YUFP}, if $A_k(\mbf{x})$ and $B_k(\mbf{x})$ are concave and convex, respectively, and the feasible set $\mathcal{X}$ is convex, then the optimization problem in~\eqref{OP:Qtrans} is concave and the optimal solution can be found by alternatively optimizing between $\mbf{x}$ and $\mbf{v}$.

We apply the above concept to~\eqref{OP1:HMS} by introducing the auxiliary variable $\mbf{v} = [v_1, v_2, \dots, v_K]$ and rewriting the objective function in~\eqref{OP1:HMS} using the quadratic transform
\ie $R = \sum_{k=1}^K \log_2(1+\gamma_k)$ where
\begin{equation}
	\gamma_k = 2v_k\sqrt{A_k(\Whms, \WD)}-v_k^2B_k(\Whms, \WD)
	\label{eq:CVXgamma}
\end{equation}
and 
\begin{equation}
	\begin{split}
		&A_k(\Whms, \WD) =\\
		&P_T T^\text{HMS}\abs{\left[\WD \Phms \Whms \right]_{k,:}\mbf{H}_{:,k}}^2
	\end{split}
\end{equation}
\begin{equation}
	\begin{split}
		&B_k(\Whms, \WD)= \\
		&\sum_{j \neq k}^K P_TT^\text{HMS}\abs{\left[\WD \Phms \Whms\right]_{k,:}\mbf{H}_{:,j}}^2 \\
		&+\sigma_\text{ant}^2T^\text{HMS} \left[\WD \Phms \Whms\right]_{k,:}\\
		&\times \mbf{\Sigma}_\text{rx}(\left[\WD \Phms \Whms\right]_{k,:})^H+\sigma_\text{RF}^2\abs{\mbf{W}^D_{k,:}}^2
	\end{split}
\end{equation}

While using~\eqref{eq:CVXgamma} renders the objective function convex in $\Whms$, it is not convex in the optimization variable, $\ahms$. On the other hand, since $\Whms = \text{diag}(\ahms)$ is a linear map of $\ahms$, the transformed objective function is a concave function of a linear map and, hence, we can find a close-to-global optimizer using algorithmic techniques such as interior-point methods~\cite[Ch.~11]{boyd2004convex}. Moreover, we recast the constraint~\eqref{OP1cnstr:pwrconservation} in quadratic form as
\begin{equation}
	(\ahms)^H \mbf{P_\text{i}}\ahms - \mbf{1}^H_N \mbf{P_\text{i}} \mbf{1}_N=0,
	\label{eq:CVXcnstr}
\end{equation}
where, $\mbf{1}_N$ is a length-$N$ vector of ones and $\mbf{P_\text{i}}$ is a diagonal matrix with its $n$-th element defined by~\eqref{eq:Pin}. 

In sum, the transformed optimization problem to optimize the achievable sum rate of HMS is 
\begin{subequations}
	\label{eq:OP2}
	\begin{align}
		\max_{\mathbf{v},\ahms, \mbf{W}^D} &\Sigma_{k=1}^K\log (1+\gamma_k)\\
		\text{s.t.} \ \ \ &  \mbf{W}^\text{HMS} = \text{diag}(\ahms)\label{OP2cnstr:Wdef}\\
		&(\ahms)^H \mbf{P}_\text{i}\ahms - \mbf{1}^H_N \mbf{P}_\text{i} \mbf{1}_N \leq 0\label{OP2cnstr:Pwrlaw}\\
		&\gamma_k = 2v_k\sqrt{A_k(\Whms, \WD)}-v_k^2B_k(\Whms, \WD)
	\end{align}	
\end{subequations}
In~\eqref{eq:OP2}, we relaxed the constraint related to the 
power conservation rule. Note that the optimum achievable sum rate occurs when the maximum signal power is coupled to the system. Consequently, at the optimal point,~\eqref{OP2cnstr:Pwrlaw} is always active. If it were not, we could rescale the optimum point to satisfy the constraint without affecting the combiner. This adjustment is possible because the radiation beam pattern of an antenna array depends on the relative phase and amplitude distribution between elements, rather than their absolute values~\cite[Ch.~6]{balanis2016antenna}. 

To solve~\eqref{eq:OP2}, we propose a two-phase algorithm, where, in the first phase, the RWD combiner\ie $\ahms$, is calculated independent of the digital combiner\ie $\WD= \mbf{I}$. By doing so, we obtain a close-to-optimum RWD combiner which focuses the received signal on the DPA by minimizing the inter-user interference. Then, we use zero-forcing (ZF) as the digital combiner to further reduce interuser interference.

To calculate the RWD combiner, as proposed in~\cite{YUFP}, we alternate between a close-to-optimum choice for $\ahms$ and $\mbf{v}$. To do so, in the first iteration, we randomly initialize $\ahms$ and scale it such that~\eqref{OP2cnstr:Pwrlaw} is satisfied with equality. Given $\ahms$ the optimum choice for the $k$-th element of $\mbf{v}$ is
\begin{equation}
	v_k^* = \frac{\sqrt{A_k(\ahms, \mbf{I}_M)}}{B_k(\ahms, \mbf{I}_M)}
\end{equation}
where, without loss of generality, we assumed the number of RF chains\ie DPA elements of HMA, is equal to number of users\ie $M=K$, and hence, $\WD$ is $\mbf{I}_M$. Then, given $\mbf{v}$, we find a close-to-optimum solution for $\ahms$ by using the interior-point method implemented in Matlab's \textit{fmincon} function~\cite{MATLAB}. The alternative optimization between $\ahms$ and $\mbf{v}$ is repeated until convergence.

Finally, the digital combiner, $\WD$ is calculated as $(\mbf{H}_\text{eq}^\text{HMS})^{-1}$ where  $\mbf{H}_\text{eq}^\text{HMS} = \mbf{G}^\text{HMS}\mbf{H}$ is the equivalent channel at the RF chain by accounting for the effect of HMS.

\textit{Single User Scenario:} Although the algorithm described above can be used for the special case of $K=M=1$, a simpler approach is a maximum ratio combiner which provides the RWD combiner in one step. Specifically, a near optimum choice for the RWD combiner is to simply compensate for the phase difference in $\mbf{H}$ and $\Phms$ to sum all the signal components coherently at the RF chain\ie
\begin{equation}
	\ahms = e^{\jmath \measuredangle (\mbf{H}^H\oslash \Phms)}
\end{equation}

\section{Power Consumption Model}\label{sec:PWR}
We were motivated to consider an HMA for its possible energy efficiency. EE is defined as
	$\EE = R/\pc$ 
where $\pc$ represents the \textit{total} power consumed by the antenna system, including its RF chains, and $R$ denotes the achievable sum rate of the antenna employed at the BS. Specifically, to fairly compare EE among different antenna designs, it is crucial to consider the various elements and driving circuitry within the antenna design. Notably, two major contributors to the power consumption of HMA are the RF chains and the circuitry needed to configure unit-cells in HMS. We begin by modeling the power consumption of the RF chains.

A simplified block diagram of a general RF chain structure is illustrated in Fig.~\ref{fig:HMA}. For a detailed version, one can refer to~\cite{AD2021RFchain}. All the elements shown in Fig.~\ref{fig:HMA} contribute to the total power consumption of the RF chain. Consequently, the power consumed by $M$ RF chains is equal
\begin{equation}
	\begin{split}
		P_{\text{RF},M} =& P_\text{OSC}+M [P_\text{filter}+P_\text{LNA}+P_\text{Mixer}+P_\text{IQD}\\
		&+2(P_\text{AAF}+P_\text{ADCdrv}+P_\text{ADC})+P_\text{CLKdist}],
	\end{split}
	\label{eq:PRF}
\end{equation}
where $P_\text{filter}$, $P_\text{LNA}$, $P_\text{Mixer}$, $P_\text{IQD}$, $P_\text{AAF}$, $P_\text{ADC}$, $P_\text{ADCdrv}$, $P_\text{OSC}$, and $P_\text{CLKdist}$ represent the power consumption of the initial filter, low noise amplifier (LNA), mixer, in-phase and quadrature demodulator (IQD), anti-aliasing filter (AAF), analog-to-digital converter (ADC), ADC driver, oscillator, and clock distribution network, respectively. Appendix~\ref{app:PCRF} details how to model and calculate the power consumption of the different elements in an RF chain. 

We next delve into modeling the power consumption of the  HMS. Each unit-cell within the HMS is controlled by applying an appropriate voltage to the implemented varactors; varactors are always in reverse bias, resulting in zero consumption power~\cite{wang2022RISPwrMeasu}. Also, the varactor provides continuous modulation of the unit-cell's weighting factor by translating the voltage applied to its terminals to the corresponding capacitive behavior. There are different ways to provide the required voltage, including using a voltage-output digital-to-analog converter (DAC)~\cite{DAC}; we adopt this approach in this paper. The power consumption associated with the DAC is denoted as $P_\text{DAC}$. Additionally, the driving data line connecting the DAC to the control FPGA and control FPGA itself contribute to the overall power consumed; these terms are represented by $P_\text{Ctrl}$ and $P_\text{FPGA}$, respectively. Notably, we assume that most of the processing occurs in the main FPGA or processor, and the HMS’ control FPGA primarily drives the varactors via the DAC, resulting in minimal power consumption for the control FPGA.

Therefore, to compute the total power consumption of the HMS, we sum the contributions from all relevant components:
\begin{equation}
	P_\text{HMS} = N(P_\text{DAC}+P_\text{Ctrl})+P_\text{FPGA}.\label{eq:PMTS}
\end{equation}
Here $N$ is the number of unit-cells in the HMS. Given the large number of unit-cells and corresponding varactors in the HMS design, accounting for all these power consumption terms is crucial for a fair comparison between different antenna designs. 

It is important to note that the power consumption of HMA is independent of its state. As described in Appendix~\ref{app:PCRF}, $P_\text{LNA}$ is the sole term in the proposed power consumption model that depends on the communication system’s state, determined particularly by the transmitted power by users $P_T$. In our analysis, we assume that all users transmit with equal power. Consequently, the combiner that optimizes the achievable sum rate would also achieve the optimum EE. We leave the problem of user power allocation to optimize the EE of an HMA-assisted system as future work.

\section{Numerical Results}\label{sec:numerical}
\rev{We now present the results of numerical simulations and compare the achievable sum rate and energy efficiency (EE) of HMAs with that of stacked intelligent metasurface antennas (SIMAs) and dynamic metasurfaces (DMAs) as MTS-based antenna designs, as well as with digital/hybrid phased arrays (DPA/HPA). We consider both single-user and multi-user uplink scenarios.
\\
\indent To ensure a fair comparison among different designs, we enforce an equal physical aperture area, ensuring that the total electromagnetic power impinging on each receiver remains the same. Additionally, we model each antenna system as a black box, capturing all inherent inefficiencies by evaluating the signal and noise power immediately after the ADC in the RF chain, where digital processing occurs. This approach allows us to assess each design’s ability to maximize SNR and, consequently, channel capacity, while accounting for real-world hardware limitations, including antenna efficiency, insertion loss of employed components, and RF chain impairments. The simulation parameters used in this evaluation are introduced next.}

\subsection	{Preliminaries}
We consider an outdoor wireless communication problem in the uplink, where $K$ users communicate with a BS equipped with either a phased array or MTS-based antenna having $N$ antennas/unit-cells. The carrier frequency, $f_c$, is set to $3$ GHz and the bandwidth to $20$ MHz. The BS is in the far field of users and, as explained in Section~\ref{sec:sysmodelPF}, the users' transmit power, $P_T$, is defined as the real part of the amplitude of the Poynting vector of their transmitted EM waves by each user.
The free space wave impedance is $\eta_0 = 377\ \Omega$ and the RF chain is assumed to be at room temperature\ie $T_\text{BS} = 290 \degree$ K. We repeat the simulation for 10 instances of user placements and 100 channel realizations per instance.

\textit{Received signal power}: The power captured by each antenna/unit-cell is equal to $P_T A_\text{eff}$ where $A_\text{eff}$ is the effective area of the antenna~\cite[Ch.~2.15.2]{balanis2016antenna}. For the MTS-based antennas, since each unit-cell is approximated by a continuous current sheet due to its subwavelength size, $A_\text{eff}$ is equal to the physical area of the unit-cell $A_{UC}$~\cite[Ch.~2.15.2]{balanis2016antenna}. For the PA, assuming an array of patch antennas, we designed a patch antenna at $f_c$ using Matlab's antenna designer~\cite{MATLAB} and calculated its $A_\text{eff}$ as given in~\cite[Ch.~2]{balanis2016antenna}.

The antenna noise power is given by $\sigma^2_{\text{ant}} = kT_EB\frac{A_\text{eff}}{\lambda_0^2}\Delta \Omega $
where $k$ is the Boltzmann constant, $T_E$ is the environment temperature which is well approximated by $290 \degree \ [K]$~\cite{Stutzman}, and $\Delta\Omega$ is the solid angle of noise source at the antenna. Assuming that the antenna noise covers half-space results in $\Delta\Omega = 2 \pi$.

\textit{Channel model}: In rich scattering environments, we use the Kronecker Rayleigh fading model where $\mbf{H} = \sigmarx\tilde{\mbf{H}}\mbf{\Sigma}_\text{tx}$ and $\mbf{\Sigma}_\text{tx}$ and $\sigmarx$ are the users' and antennas' correlation matrices, respectively, and $\tilde{\mbf{H}} \sim \mathcal{CN}(\mbf{0},\mbf{I}_{NK})$. We assume users are distributed uniformly in a hexagon of radius 250 m within an exclusion zone of 25 m around the BS and the channel between them is uncorrelated hence, $\mbf{\Sigma}_\text{tx} = \text{diag}(\beta_1, \dots, \beta_K)$ where $\beta_k$ is the path loss of the $k$-th user, computed using the Winner II urban microcell non-line of sight scenario~\cite{Winner}. Assuming scatterers are distributed in 3D space, the antenna's correlation matrix is defined as $(\sigmarx)_{n,n'} = \sinc\left(\frac{2d_{n,n'}}{\lambda_0}\right)$~\cite{paulraj2003introduction} where $d_{n,n'}$ is the distance between the $n$-th and $n'$-th unit-cell. In both rich scattering and realistic 3GPP channels~\cite{quadriga}, the external noise correlation matrix is equivalent to the antenna correlation matrix in a rich scattering environment\ie $\zant\sim \mathcal(0,\sigma^2_{\text{ant}} \sigmarx )$.

\subsection{Comparison with other designs} 
To facilitate a meaningful comparison between HMA and other antenna configurations in terms of achievable sum rate and EE, this section thoroughly examines the system model for each antenna design \rev{accounting for the inefficiencies of their integrated components}. Additionally, we emphasize the algorithms utilized to optimize the achievable sum rate across various antenna designs, while also considering their associated power consumption.

\subsubsection{Digital Phased Array}
In DPAs with $N$ antennas, each antenna is connected to an RF chain and a digital combiner, $\mbf{G}^D$, is employed. Hence, the processed signal $\mbf{y}$ at the antenna is described as $\mbf{y}= g\mbf{G}^D(\mbf{x}+\zrf) $. To estimate the achievable sum rate in these systems, we employ maximum ratio combiner (MRC) and zero-forcing combiner (ZFC) for single-user and multi-user scenarios, respectively. Finally, the power consumed by the digital MIMO array is equal to the power consumption of $N$ RF chains.

\subsubsection{Hybrid Phased Array}
\begin{figure}[t]
	\begin{subfigure}{\linewidth}
		\centering
		\includegraphics[width=0.72\linewidth]{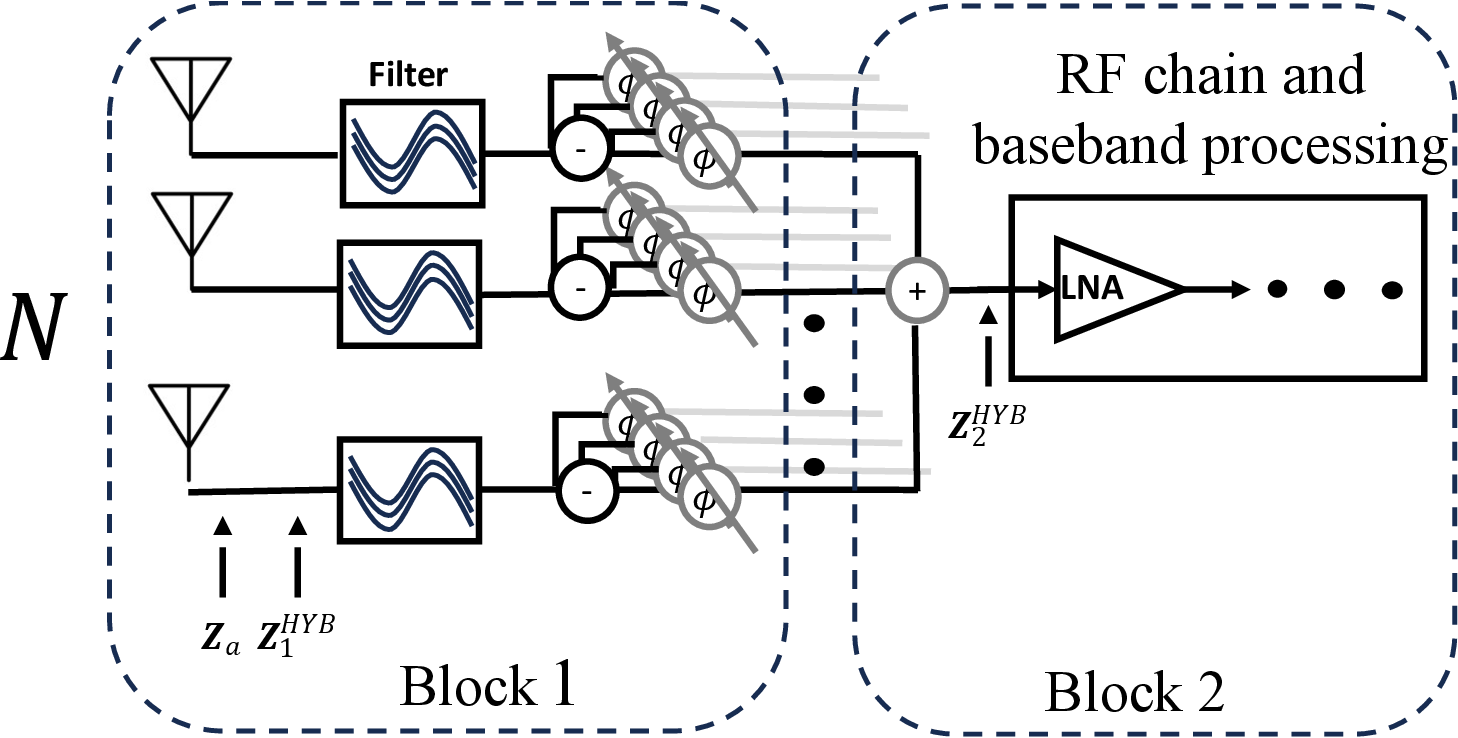}
		\caption{}
		\label{fig:HYBfull}
	\end{subfigure}
	\hfill
	\begin{subfigure}{\linewidth}
		\centering
		\includegraphics[width=0.72\linewidth]{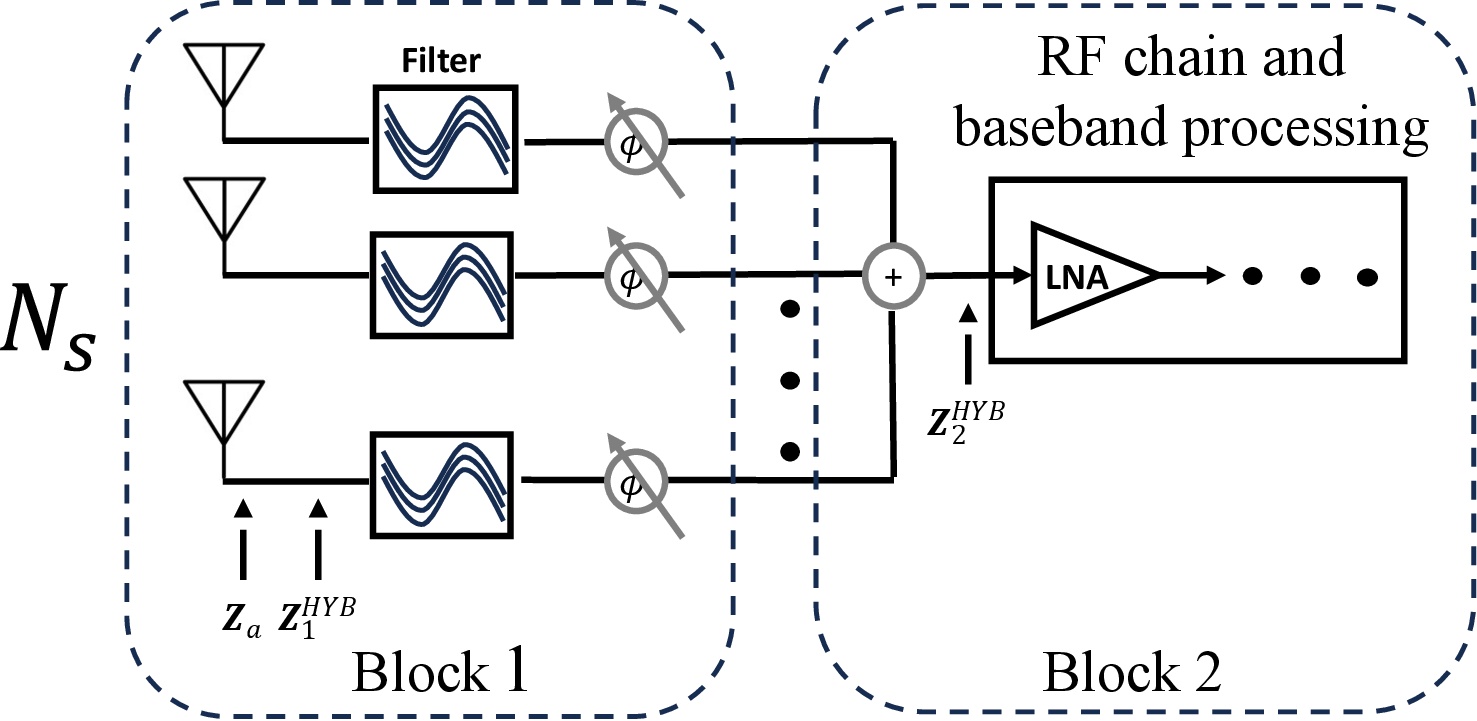} 
		\caption{}
		\label{fig:HYBpartial}
	\end{subfigure}
	\caption{Hybrid MIMO architecture and its relative placement with respect to RF chain components (a) Partially-connected hybrid MIMO, (b)  Fully-connected hybrid MIMO }
	\label{fig:HYB}
\end{figure}
The two main configurations of HPA are fully-connected and partially-connected analog combiners. In a fully-connected HPA (FCHP), as shown in Fig.~\ref{fig:HYBfull}, $N$ antennas are connected to $M$ RF chains by using $MN$ phase-shifters (PSs), an $M$-way power divider, and an $N$-way power combiner as analog combiners. In contrast, as depicted in Fig.~\ref{fig:HYBpartial}, a partially-connected HPA (PCHP), connects a sub-array of $N_s$ antennas to each RF chain using $N$ PSs and an $N_s$-way power combiner. The deployed PSs in both structures can be passive (PAS) or active (ACT). When using passive PSs, the insertion loss, $L_\text{PS}$, of the PS is imposed on the signal. On the other hand, deploying active PSs provides the opportunity to modulate the amplitude and phase of the signal independently using a vector modulator (VM) followed by an attenuator~\cite{ActivePS,ADVM} at the cost of adding to the power consumption of the HPA.

For simplicity of analysis, we break the structure of HPA in two blocks as shown in Fig.~\ref{fig:HYB}. The added noise by the first block is colored due to passing through the analog combiner, while the added noise by the second block remains white. Therefore, the system model for a hybrid MIMO system is:
\begin{equation}
	\mbf{y} = g_2^\text{HYB}\mbf{W}^D(g_1^\text{HYB}\mbf{W}^\text{ABF}(\mbf{x}+\zant+\mbf{z}_{1}^\text{HYB})+\mbf{z}_2^\text{HYB})
	\label{eq:HYBs}
\end{equation}
where $\mbf{W}^\text{ABF} \in \mathbb{C}^{M \times K}$ represents the analog combiner, and $g_i^\text{HYB}$ and $\mbf{z}_{i}^\text{HYB}$ are the gain and noise of the $i$-th block, $i \in \{1,2\}$, in HPA structure, respectively.

\textit{Hardware limitations}: In the HPA structure, the signal losses happen in PSs and the Wilkinson power divider/combiner represented by $L_\text{PS}$ and $L_\text{D,C}$, respectively. Similar to~\cite{RiberioHYBInsertionloss}, we assume an $x$-way power divider/combiner is formed by cascading $\lceil \log_2(x) \rceil$ 2-way power divider/combiner. Hence, the insertion loss of the divider/combiner would be $L_\text{D,C} = \lceil \log_2(x) \rceil L_\text{2\ D,C}$ where $L_\text{2\ D,C}$ is the insertion loss of a 2-way power divider/combiner. These values directly impact the gain of the first and second blocks in the HPA structure, represented by $g_1^\text{HYB}$ and $g_2^\text{HYB}$, respectively.

\textit{Achievable sum rate}: In the case of PCHP and FCHP with passive PSs, we use algorithms proposed in~\cite{MOHYB} and~\cite{SohrabiHYB}. Moreover, when active PSs are employed in a FCHP, we use MRC and ZFC as in a DPA and scale the amplitude of combiner elements to be less than one to match the requirements of an active analog combiner. In the case of a PCHP with active PSs, we use the algorithm proposed in~\cite{MOHYB}.

\textit{Power Consumption}: An HPA compromises $M$ RF chains and $N_p$ PSs, therefore, $P_\text{HPA} = P_{\text{RF},M}+N_pP_\text{PS}$.

\subsubsection{Other MTS-based antenna designs}
We assume a 5-layer SIMA which achieves the highest sum rate when accounting for the number of layers~\cite{SIMuplinkEfficiency}. Also, the achievable sum-rate of DMAs and SIMAs is calculated using non-zero mapper (NZM) and closed-form mapper (CFM) as proposed in~\cite{DMAuplinkEfficiency} and the IPM and gradient ascend (GA) algorithms proposed in~\cite{SIMuplinkEfficiency}, respectively. Also, the power consumption of these designs is calculated using the method explained in Section~\ref{sec:PWR} since they are MTS-based antenna designs with the same hardware requirements. For completeness a summary of the reported techniques in~\cite{SIMuplinkEfficiency} and~\cite{DMAuplinkEfficiency} is mentioned in \rev{Supplement materials}.
\begin{table}[t]
	\captionof{table}{Simulation Parameters \label{table:simParam}}
	\begin{subtable}{\linewidth}
		\centering
		\begin{tabular}{|c|c|c|c|c|}
			\hline 
			\rowcolor{gray!50}
			\cellcolor{gray!100}
			& {PA} & DMA& HMS&SIM\\
			\hline
			\thead{Insertion\\loss} & \thead{0.9~\cite{Patchantenna}}& 0.1~\cite{DMAuplinkEfficiency} & 0.7~\cite{ataloglou2023metasurfaces}&0.7~\cite{phaseonlyTransmissiveMTS}\\
			\hline
			Thickness & - & \thead{1.52 [mm]~\cite{DMAsub6Uplink}} & $\lambda_0$& $5\lambda_0$\\
			\hline
			\thead{Unit-cell\\size}& \small \thead{$\frac{\lambda_0}{2}\cdot \frac{\lambda_0}{2}$}& \small \thead{$\frac{\lambda_0}{2}\cdot \frac{\lambda_0}{6}$~\cite{SmithTermination2018dynamically}}& \small \thead{$\frac{\lambda_0}{4}\cdot \frac{\lambda_0}{4}$}& \small \thead{$\frac{\lambda_0}{2}\cdot \frac{\lambda_0}{2}$~\cite{SIM_ICC}}\\
			\hline
			$A_\text{eff}$& \small \thead{$0.0026$}& \small $0.00084$& \small $0.000625$& \small $0.0025$\\
			\hline
		\end{tabular}
		\caption{Antenna design related parameters}
		\centering
		\begin{tabular}{|c|c|c|c|}
			\hline
			\rowcolor{gray!50}
			{Device }& Power [W] & Gain [dB] & NF [dB]\\
			\hline
			{Filter~\cite{LPF} }& 0 & -3 & 3\\
			\hline
			{LNA~\cite{LNA} }& 0.75& 15 & 3.5\\
			\hline
			{Mixer~\cite{mixer} }& 0.4 & -7.1 & 8.0\\
			\hline
			{IQD~\cite{IQdem}}& 2.2 & 0 & 31\\
			\hline
			{AAF~\cite{AAF}}& 0.01 & 0 & 0\\
			\hline
			{ADC driver~\cite{ADCdrv}}& 0.15 & 20 & 6.4\\
			\hline
			{ADC~\cite{ADC}}& 0.725 & - &  30\tablefootnote{Calculated using information provided in~\cite{ADCNF} and signal-to-noise-and-distortion ratio of ADC}\\
			\hline
			{OSC~\cite{CLK}}& 0.02& -& -\\
			\hline
			{Clock distribution~\cite{CLKdst}}& 0.08&-&-\\
			\hline
			{PAS PS~\cite{PSpassive}}& 0 & -5& 5\\
			\hline
			{ACT PS~\cite{VM}}& 0.625& -4.5 & 23\\
			\hline
			{Wilkinson~\cite{wilkinson}} & 0 & -3.9 & 3.9\\
			\hline
			{DAC~\cite{DAC}}& 0.002\tablefootnote{per channel}& -&-\\
			\hline
			{HMS/MTS controller\tablefootnote{Calculated using  Xilinx Power Estimator (XPE)~\cite{AMDxilinx} for one low voltage differential signaling line per DAC and on a per channel basis}}& 0.0006& -&-\\
			\hline
			{FPGA\tablefootnote{Calculated using XPE}}& 0.1&-&-\\
			\hline
		\end{tabular}
		\caption{Hardware-related parameters }

	\end{subtable}
	\label{table:simparam}
\end{table}

Finally, the noise figure (NF) and gain of the RF chain and different blocks of HPA is calculated using the technique proposed in~\cite{CascadedNF, NFparallel}. A summary of the hardware specific simulation parameters is provided in Table~\ref{table:simParam} where the noise figure of a passive element is set equal to its insertion loss~\cite{bhattacharyya2006phased}. Also, the efficiency of DMA reported in Table~\ref{table:simparam} is equal to the maximum weighting factor of each unit-cell in DMA~\cite{DMAuplinkEfficiency}.

\subsection{Simulation Results}
\subsubsection{Power consumption} 
\begin{figure}[t!]
	\centering
	\includegraphics[width=\linewidth]{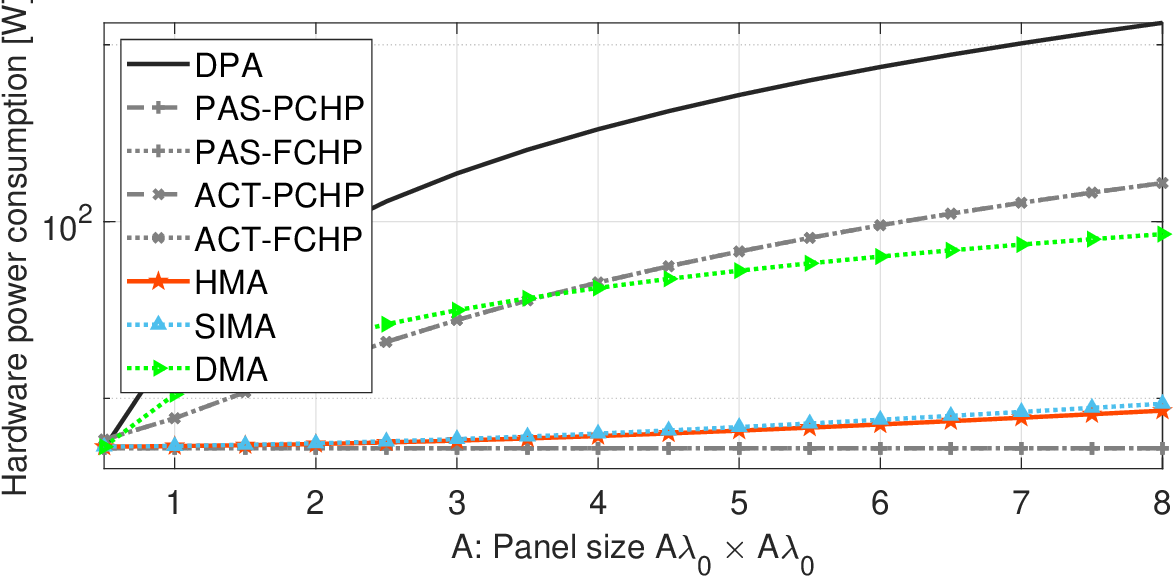}
	\caption{Power consumption vs. physical aperture area, $P_T = 20\ [\frac{dBm}{m^2}]$. 
	}
	\label{fig:PwrconsVsA}
\end{figure}

To show the effectiveness of an HMA as the antenna choice for the next generation of wireless communications, we begin by comparing the power consumption of different designs when the number of RF chains in the HPA, SIMA, and HMA are set to $M=1$. Due to their structure, $M$ in a DMA and DPA is a function of aperture area and, hence, increases with antenna size. As depicted in Fig.~\ref{fig:PwrconsVsA}, HMAs and SIMAs exhibit minimal power increase as the antenna size grows, thanks to their fixed RF chain count. In contrast, a DPA consumes the most power due to its large number of RF chains. Additionally, an HMA’s power consumption is slightly lower than a SIMA’s, attributed to their structural differences and fewer unit-cells in an HMA.

\begin{figure*}[!t]
	\begin{subfigure}{0.328\textwidth}
			\centering
			\includegraphics[width=\linewidth]{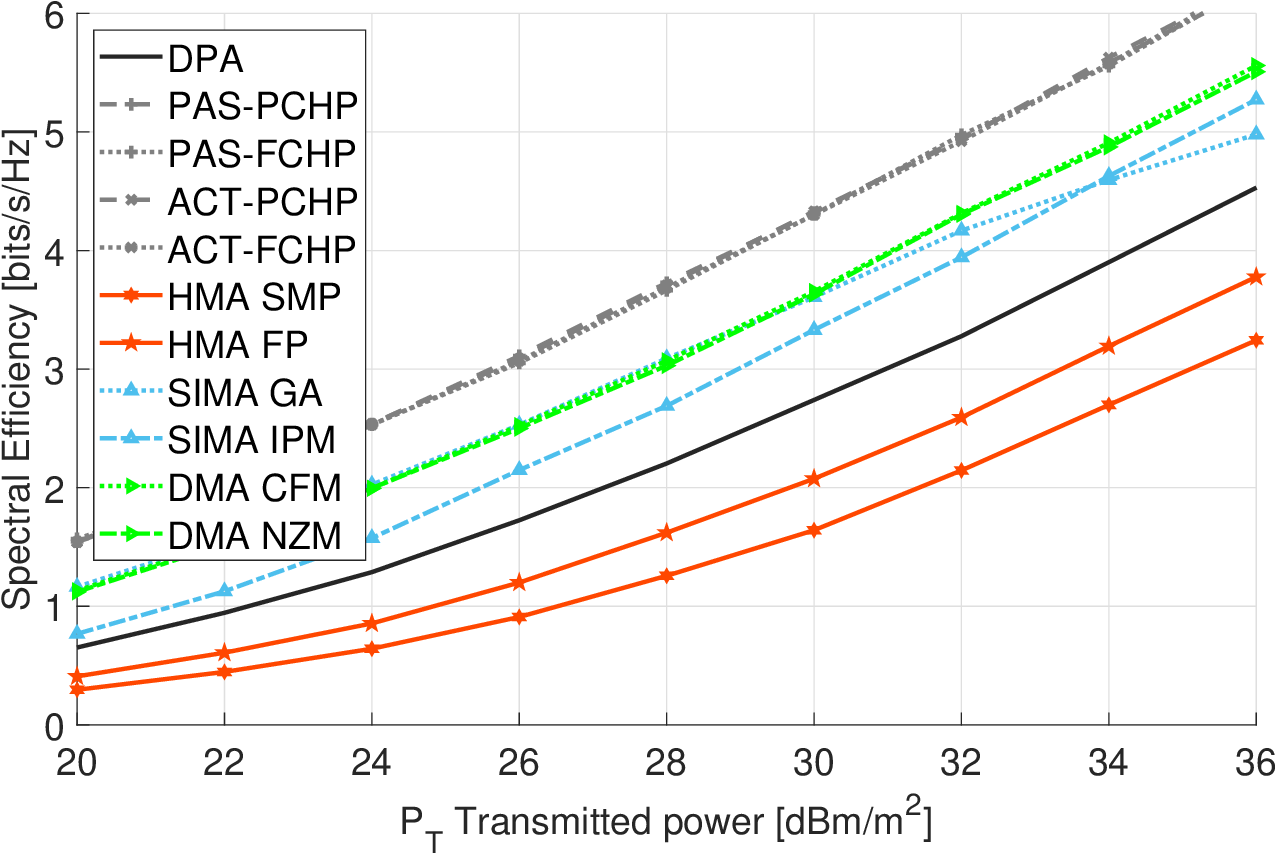}
			\caption{}
			\label{fig:NolossSESU}
	\end{subfigure}
	\hfill
	\begin{subfigure}{0.328\textwidth}
			\centering
			\includegraphics[width=\linewidth]{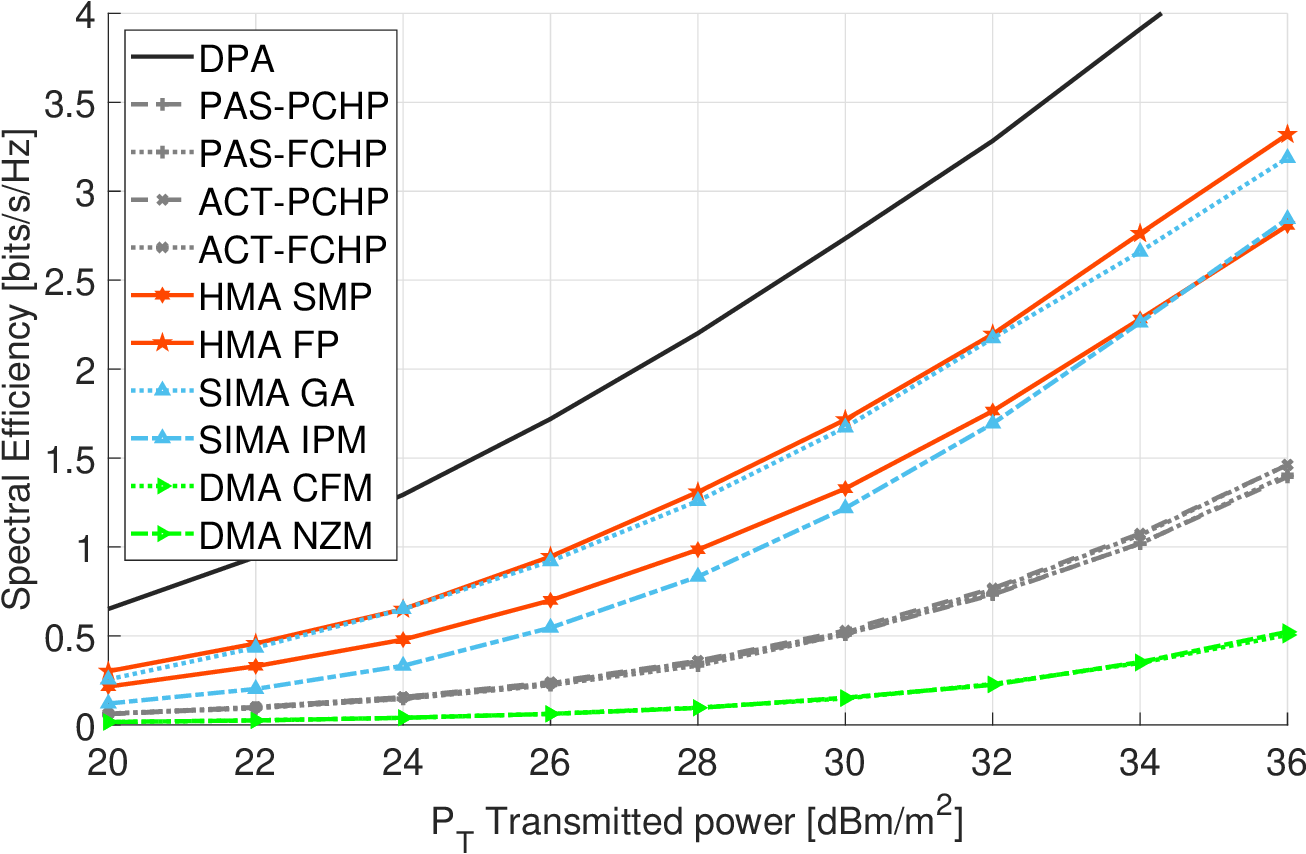}
			\caption{}
			\label{fig:SERSU}
	\end{subfigure}
	\hfill
	\begin{subfigure}{0.328\textwidth}
			\centering
			\includegraphics[width=\linewidth]{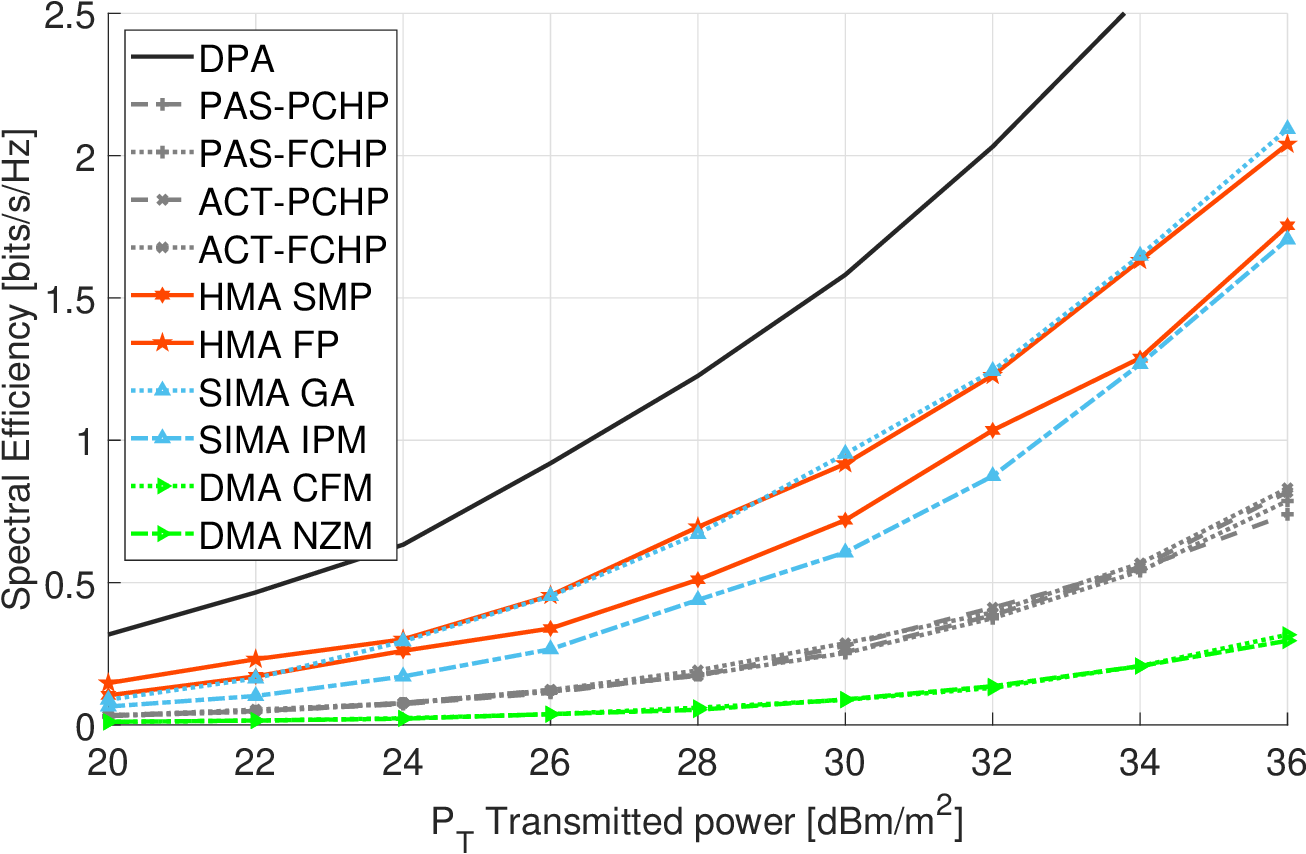} 
			\caption{}
			\label{fig:SEQSU}
	\end{subfigure}
	\caption{Achievable sum rate of different antenna designs in (a) Not considering hardware losses and considering hardware impairments in (b) rich scattering (c) realistic 3GPP channels under equal physical aperture area. }
	\label{fig:SESU}
\end{figure*}

\begin{figure}[!t]
	\centering
	\begin{subfigure}{0.47\linewidth}
			\centering
			\includegraphics[width=\linewidth]{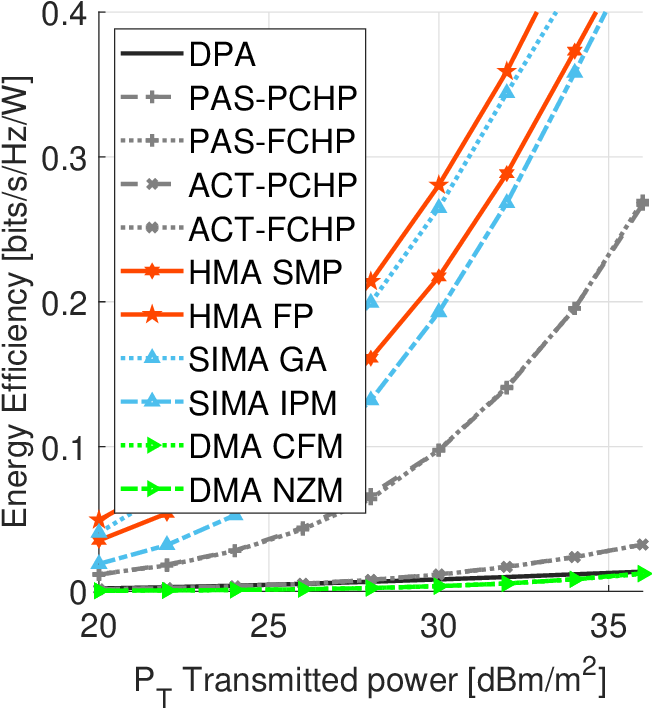}
			\caption{}
			\label{fig:EERSU}
	\end{subfigure}
	\hfill
	\begin{subfigure}{0.47\linewidth}
			\centering
			\includegraphics[width=\linewidth]{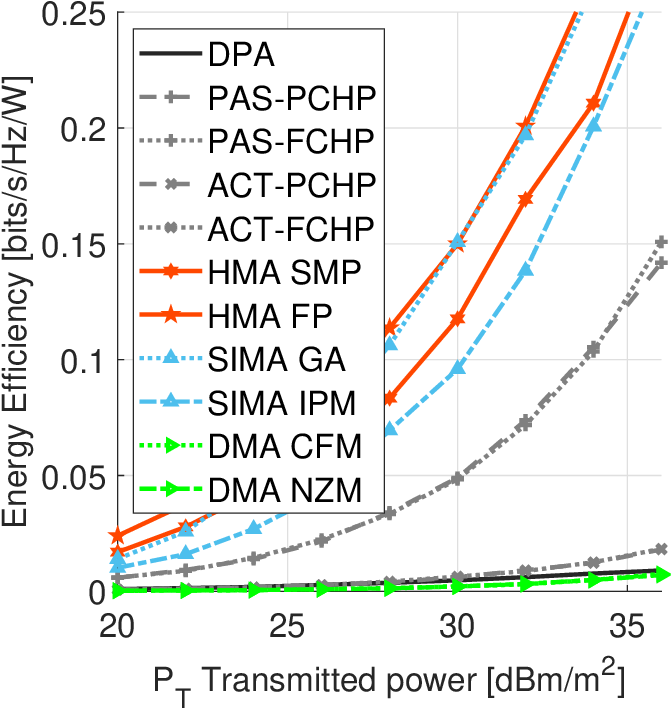} 
			\caption{}
			\label{fig:EEQSU}
	\end{subfigure}
	\caption{Energy efficiency of different antenna designs in (a) rich scattering (b) realistic 3GPP channels under equal physical aperture area. }
	\label{fig:EESU}
\end{figure}

In the following, we compare the achievable sum rate and EE of HMA with other antenna designs. The HMA performance using the RWD combiner calculated by the FP-based algorithm as proposed in Section~\ref{sec:WDFP} is labeled with ``FP'' while the RWD combiner calculated by the simple algorithm is labeled as ``SMP'' in Figs.~\ref{fig:SESU} - \ref{fig:EEMU}. Also, we limit the number of iterations of the IPM to 100 in the first round of the FP-based algorithm and 20 in the successive rounds.

\subsubsection{Single-user Scenario}
Depicted in Fig.~\ref{fig:SESU} is the achievable sum rate of different designs with an aperture area of $4\lambda_0 \times 4\lambda_0$ with $K=1$. We set $M=1$ for the HMA, SIMA, and HPA while, dictated by the structure of the DMA and DPA, $M$ in these designs is $8$ and $64$, respectively. To show the importance of accounting for hardware limitations when comparing different designs, we also report in Fig.~\ref{fig:NolossSESU} the achievable sum rate of different systems in rich scattering channels assuming perfect hardware. We do this by setting the efficiency parameter of all designs to one and insertion losses of analog combiners in HPA to zero. In this case, the HPA, DMA, and SIMA achieve the highest sum rates. This is mainly due to the limited number of RF chains in their systems. Without inter-user interference (since $K=1$), a smaller value of $M$ results in lower noise power and therefore, higher signal-to-noise ratio (SNR). 

When accounting for hardware limitations, as depicted in Fig.~\ref{fig:SERSU} and Fig.~\ref{fig:SEQSU} for rich scattering and realistic 3GPP channels, respectively, the achievable sum rate of an HMA using the FP algorithm surpasses that of other antenna designs. Importantly, the HMA using the SMP algorithm achieves the same level of performance as the 5-layer SIMA using IPM. Also, the achievable sum rate of the HMA is close to that of a DPA illustrating the effectiveness of the RWD combiner in the HMS in focusing the received signal on the single RF chain.

Furthermore, illustrated in~\cref{fig:EESU} is the EE of each antenna design in both rich scattering and realistic 3GPP channels. As depicted, in both scenarios, HMA has the best EE followed by the SIMA and PAS-HPAs. This is directly due to lower power consumption and higher achievable sum rate of HMA compared with other antenna designs.

\subsubsection{Multi-user Scenario}
\begin{figure*}[!t]
	\begin{subfigure}{0.328\textwidth}
			\centering
			\includegraphics[width=\linewidth]{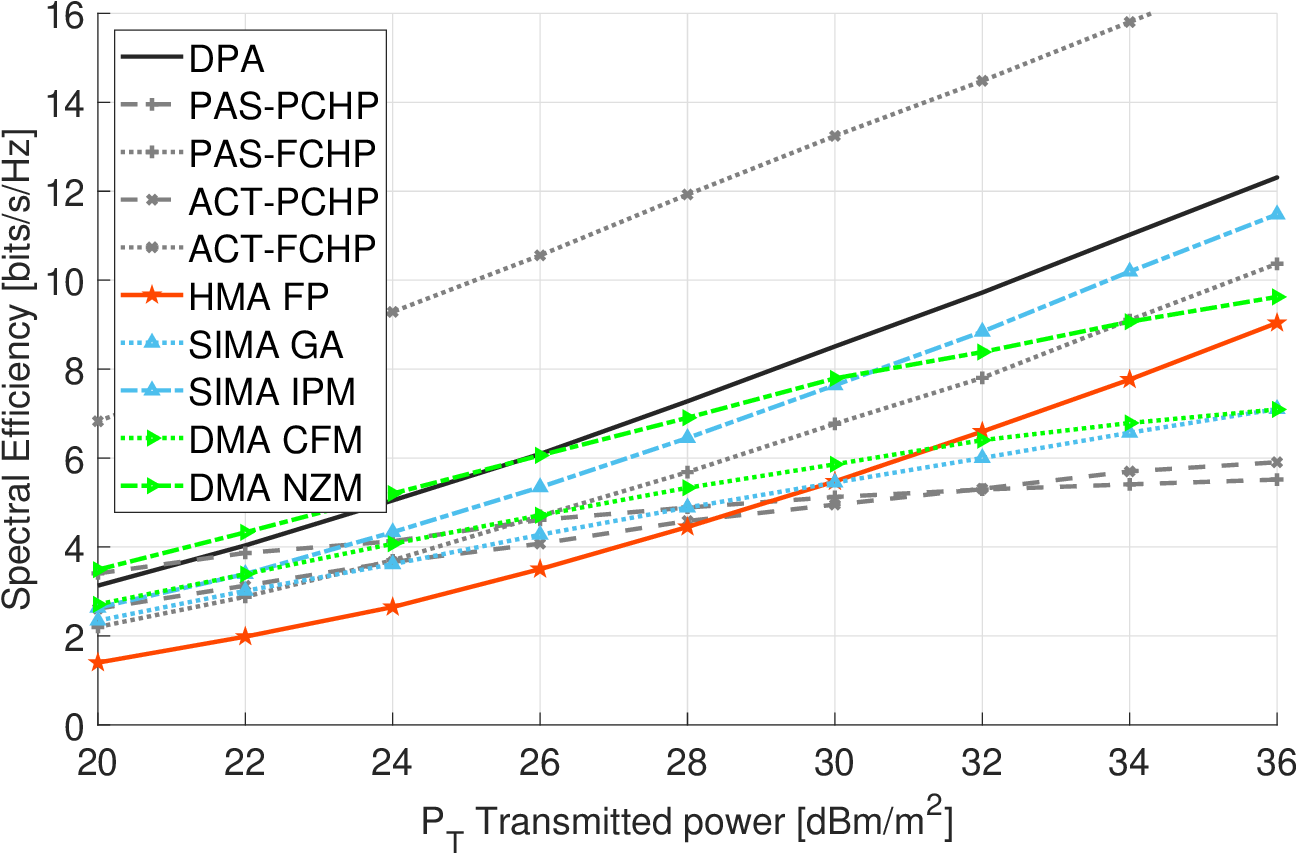} 
			\caption{}
			\label{fig:NolossSEMU}
	\end{subfigure}
	\hfill
	\begin{subfigure}{0.328\textwidth}
			\centering
			\includegraphics[width=\linewidth]{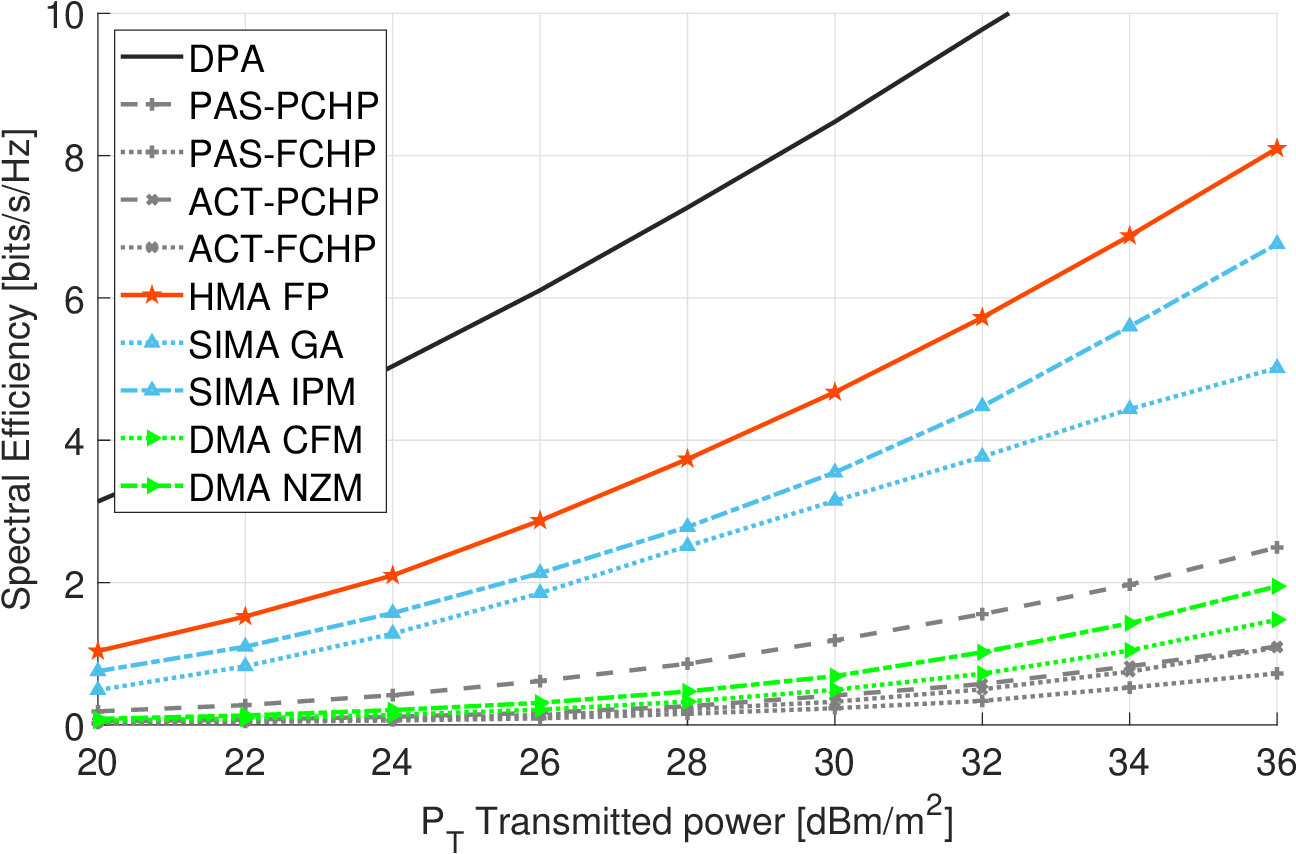}
			\caption{}
			\label{fig:SERMU}
	\end{subfigure}
	\hfill
	\begin{subfigure}{0.328\textwidth}
			\centering
			\includegraphics[width=\linewidth]{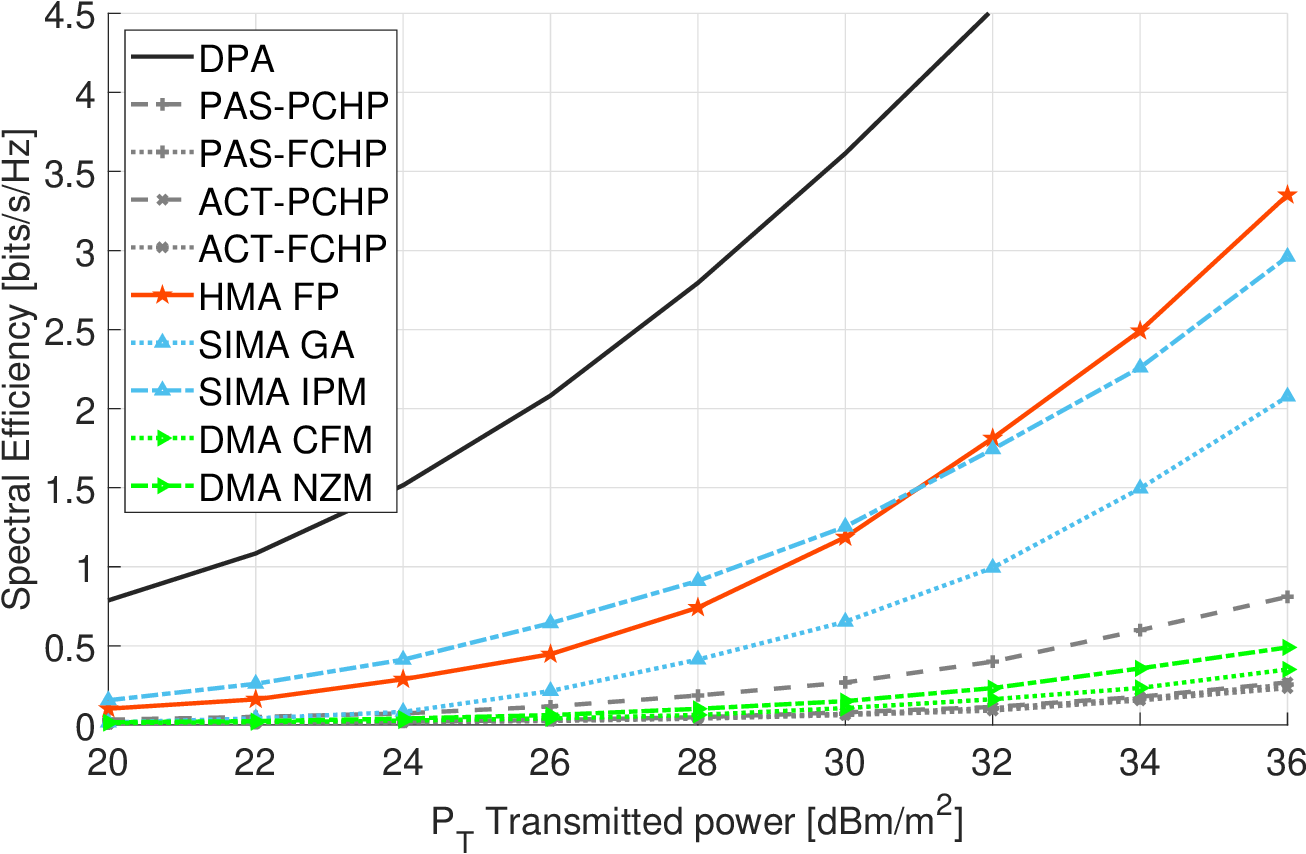} 
			\caption{}
			\label{fig:SEQMU}
	\end{subfigure}
	\caption{Achievable sum rate of different antenna designs in (a) Not considering hardware losses and considering hardware impairments in (b) rich scattering (c) realistic 3GPP channels under equal physical aperture area. }
	\label{fig:SEMU}
\end{figure*}

In the multi-user scenario, we assume the BS serves $K=2$ randomly placed users. \rev{This provides a fundamental performance benchmark while maintaining analytical tractability. This setup allows us to isolate and compare the impact of different antenna designs on mitigating inter-user interference and optimizing the achievable sum rate. While real-world communication systems often involve a larger number of users, adding more users introduces additional system-level considerations such as power control and scheduling. Therefore, to focus on the core hardware and signal processing capabilities of each antenna design, we adopt a two-user setup with simultaneous transmission and equal power allocation. It is important to note that the proposed framework and optimization approach are generalizable and can be extended to scenarios with more users. Furthermore,} the number of RF chains in the HMA, SIMA, and HPA are set equal to the number of users\ie $M=2$. Fig.~\ref{fig:NolossSEMU} illustrates the achievable sum rate in rich scattering channels using ideal hardware. In this impractical case, due to inter-user interference, the achievable sum rate of PCHPs, both passive and active, and DMA saturates as $P_T$ increases. This is while the performance of the SIMA, FCHPs, and HMA do not saturate, i.e., show a continuous increase in achievable sum rate as $P_T$ increases. The best performance is demonstrated by FCHP with passive PSs where the noise power is the lowest. 

On the other hand, as depicted in Fig.~\ref{fig:SERMU}, when accounting for hardware inefficiencies, the HMA achieves the highest sum rate in rich scattering channels. This demonstrates the importance of independent control over phase and amplitude of the weighting factor of each unit-cell to suppress inter-user interference. Furthermore, as illustrated in Fig.~\ref{fig:SEQMU}, in realistic 3GPP channels, the achievable sum-rate of the HMA and SIMA are very close to each other. For high $P_T$s, higher than $31$ dBm, the HMA's performance surpasses that of the SIMA. As the interference between users is the dominant term in reducing the achievable sum rate of different designs, the DPA offers the highest rates by effectively canceling the inter-user interference.

\begin{figure}[!t]
	\centering
	\begin{subfigure}{0.47\linewidth}
			\centering
			\includegraphics[width=\linewidth]{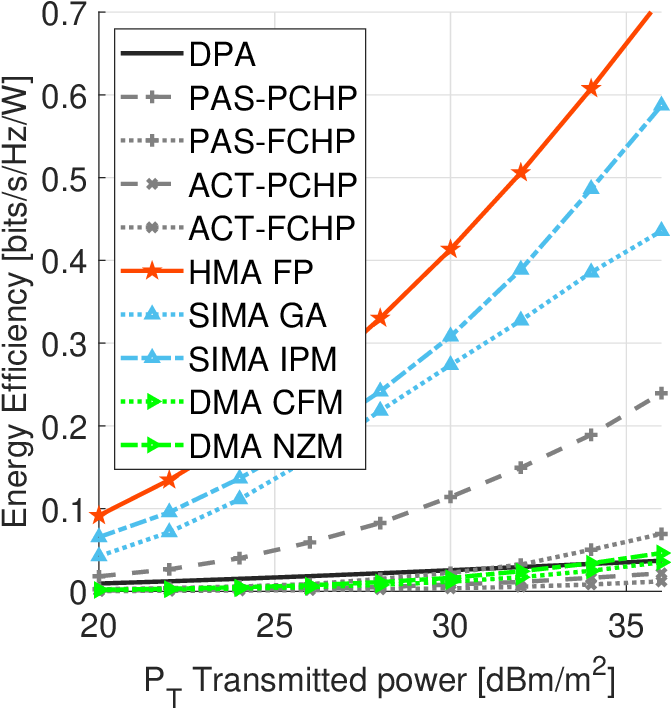}
			\caption{}
			\label{fig:EERMU}
	\end{subfigure}
	\hfill
	\begin{subfigure}{0.47\linewidth}
			\centering
			\includegraphics[width=\linewidth]{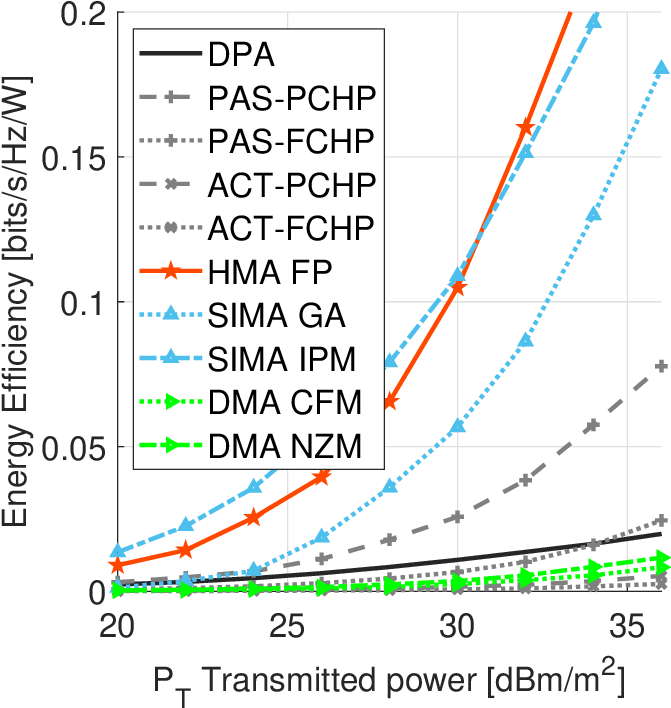} 
			\caption{}
			\label{fig:EEQMU}
	\end{subfigure}
	\caption{Energy efficiency of different antenna designs in (a) rich scattering (b) realistic 3GPP channels under equal physical aperture area. }
	\label{fig:EEMU}
\end{figure}

Finally, depicted in Fig.~\ref{fig:EEMU}, is the EE for different antenna designs in rich scattering and realistic 3GPP channels. As shown, the HMA has the highest efficiency among the different designs considered due to its high achievable sum rate and low power consumption. After the HMA, the most efficient designs are the SIMA and PCHP with active PSs.

\subsubsection{A Note on Hardware Complexity and Processing Time}
For the next generation of wireless communication, the antenna designs should demonstrate low hardware implementation complexity and low usage of natural resources~\cite{6Gvision}. While both HMA's and SIMA's achievable sum rates are comparable in various scenarios, HMA's hardware complexity is significantly lower. The HMA is formed by one layer of a HMS where each unit-cell is formed by stacking up to four impedance sheets. This is while a SIMA is formed by $L$ layers; $L=5$ show the same performance as a HMA, where each MTS layer is formed by stacking three impedance sheets to provide full control over phase while keeping the amplitude of weighting factor constant~\cite{SIMuplinkEfficiency}. The simplicity of an HMA, in terms of hardware complexity, makes an HMA the better choice, as compared with a SIMA, where the rate performance of both is at an acceptable level. Table~\ref{table:HWcompl} provides an overview of the hardware complexity for each antenna design.

Furthermore, one deciding factor in choosing an antenna design for the next generation of wireless communications is the complexity of the required signal processing. A DPA only needs the calculation of MRC or ZFC matrix in single-user and multi-user scenarios. 
\rev{On the other hand, HPA, DMA, SIMA, and HMA all require algorithmic techniques that demand a large number of iterations to converge, adding to both the processing delay and the complexity of the employed signal processing techniques~\cite{Hint}. \rev{Since iterative algorithms inherently limit parallelization, execution time is primarily dictated by the number of iterations required for convergence\footnote{Other factors that affect the execution delay of an algorithm implemented on an FPGA include algorithmic-level parallelization, memory organization, and computational complexity~\cite{Processingtime2018survey}.}}.  Although MTS-based antennas (i.e., HMA, SIMA, and DMA) enable part of the processing in the radiated wave domain at the speed of light, determining the optimal MTS configuration still necessitates iterative computations, introducing additional delays that scale with the iteration count.}

\begin{table}[t]
	\centering
	\begin{tblr}{row{1} = {bg=lightgray}, cells={valign=m,halign=c},colspec={QQQQQQ}, rowsep = 0pt,hlines,vlines}
		Design & Antennas & \thead{RF\\chains} & \thead{MTS\\layers} & \thead{Unit-\\cell} & PS\\
		\thead{DPA} & $N$ & $N$ & - & - & -\\
		HPA & $N$ & $K$ & - & - & \thead{FC: $KN$\\PC: $N$}\\
		HMA & $K$ & $K$ & $1$ & \thead{4-layer\\stacked} & -\\
		SIMA & $K$ & $K$ & $L>1$ & \thead{3-layer\\stacked} & -\\
		DMA & - & $\lceil 2A_z/\lambda \rceil$  & $1$ & \thead{Single\\layer} & -\\
	\end{tblr}
	\caption{Hardware complexity of each antenna design of size $A_z \times A_x$ when serving $K$ users.}
	\label{table:HWcompl}
\end{table}

For each of these designs to play a role in the next generation of wireless communications a deciding factor is the required signal processing techniques to be effective and simple. In the case of serving a single user, we have seen that the HMA offers a simple solution with performance comparable to that of other designs. Also, due to the nature of the weighting factors in a HMS, the HMA's optimization problem\ie\eqref{eq:OP2}, is closest to a convex problem among all MTS-based antenna designs, which results in finding a close-to-optimum yet simple solution relatively straightforward. In summary, we believe we have shown the great potential of HMA as the antenna of the next generation of wireless communications and leave proposing a low-complexity technique optimizing the achievable sum-rate for future work.

\section{Conclusions}\label{sec:conclusion}
\rev{In this paper, \revtwo{we propose the use of Huygens’ metasurface-based antennas (HMAs) as the primary choice for next-generation wireless communications}. Our analysis to support this contention integrates both electromagnetic theory (EM) and information theory (IT), providing a comprehensive understanding of HMA performance. We introduce a simple yet effective model that captures various aspects of an HMA and its electromagnetic properties, including its hardware limitations. We then formulate an optimization problem to maximize the achievable sum rate in these systems, combining IT and EM theory to address the unique properties of HMA-assisted wireless communication systems.} To assess the energy efficiency of the system, we develop a power consumption model that considers both the RF chains and the driving circuitry of the HMA.
We compare the achievable sum rate and energy efficiency of an HMA with that of a DPA, HPA, DMA, and SIMA, accounting for the hardware limitations of each design. For a fair comparison, we ensure all designs have an equal aperture area constraint. Importantly, our results demonstrate the significant potential of HMAs in offering energy-efficient, high-throughput solutions for wireless networks.

\rev{While this paper highlights the potential of HMAs in improving energy efficiency and throughput in wireless networks, there are several practical challenges inherent to all MTS-based antenna designs. These challenges include the limited operating bandwidth of MTSs, constrained by the resonance nature of the unit-cells, insertion losses of the MTS dominated by dielectric losses and dissipation in tunable components~\cite{ataloglou2023metasurfaces}, and the need for low-complexity, low-delay algorithms to compute the optimal state of the MTS. Addressing these issues is crucial for realizing the full potential of HMAs in real-world systems, and will require further research and development.}



\appendices

\section{Power Consumption of RF chain} \label{app:PCRF}
To calculate the power consumption of the RF chain, we directly extract the reported power consumption in the corresponding device's datasheet. The exception is the LNA, where its power consumption model is described below. In this appendix, we also provide a more detailed explanation of the employed ADC's power consumption model.

The power consumption of LNA has two components. The first is the static power consumption, $P_\text{stat}$, and the second depends on the output signal power and the power added efficiency (PAE) of the LNA~\cite{PAE}. PAE is defined as $\text{PAE}=\eta=\frac{P_\text{out}-P_\text{in}}{P_\text{LNA}}$; when $P_\text{out} = G_\text{LNA}P_\text{in}$, where $G_\text{LNA}$ is LNA's gain, then $P_\text{LNA} = \frac{G_\text{LNA}-1}{\eta}P_\text{in}$. 

The PAE is usually measured at the saturation when the LNA efficiency is at its maximum. When the signal power $P_\text{in}$ is very small, power consumption is dominated by the static term. Hence, we model $P_\text{LNA}$ as $\max (P_\text{stat}, \frac{G_\text{LNA}-1}{\eta}P_\text{in})$. In this paper, we set PAE to $12\%$~\cite{LNA}. We note that in the case of MTS-based antennas and HPAs where multiple antennas are connected to a single RF chain, $P_\text{in}$ of the LNA is equal to sum of the signal power captured by all the connected antennas. Also, we calculate the power captured by each antenna in an on-average basis where the power reaching each antenna is approximated by the average path loss of the users multiplied by $A_\text{eff}$ of each antenna, and $P_T$ of the users.

The common practice to calculate the power consumption of an ADC is to assume that $P_\text{ADC}$ is linearly proportional to the sampling frequency of ADC and exponentially proportional to its number of bits~\cite{ADCPwrConsFlash}; this is the case for a flash ADC famous for its fast response time and high power consumption. Modern BSs use pipelined ADCs~\cite{pipelinedADCmain} which offer low power consumption simultaneously with a short response time. A detailed model for calculating the power consumption of a pipelined ADC is provided in~\cite[Ch.~4.4.1]{ADCplPwerModel}, which heavily depends on the ADC's internal structure. Hence, we assume employing the same ADC\eg~\cite{ADC}, across different antenna designs, in terms of the number of bits and sampling frequency and use the power consumption as reported in the chosen ADC's datasheet.


\bibliographystyle{IEEEtran}
\bibliography{references}

 

 \begin{IEEEbiography}[{\includegraphics[width=1in,height=1.25in,clip,keepaspectratio]{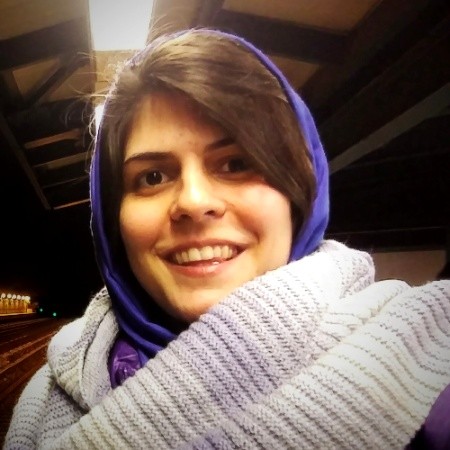}}]{Maryam Rezvani}
 (Student member IEEE) received a B.Sc. degree in electrical engineering (majored in communication systems) in 2012 from the Sharif University of Technology and an M.Sc degree in electrical engineering (majored in electromagnetic fields and waves) in 2015 from the University of Tehran, Tehran. She worked for about ten years as a senior system designer in the network industry. She is currently a Ph.D. student working with Prof. Adve at the ECE department, University of Toronto, where Her research centers around efficient MIMO processing, holographic MIMO, and electromagnetic information theory.
 \end{IEEEbiography}
 \vspace{1pt}
 \begin{IEEEbiography}
 [{\includegraphics[width=1in,height=1.25in,clip,keepaspectratio]{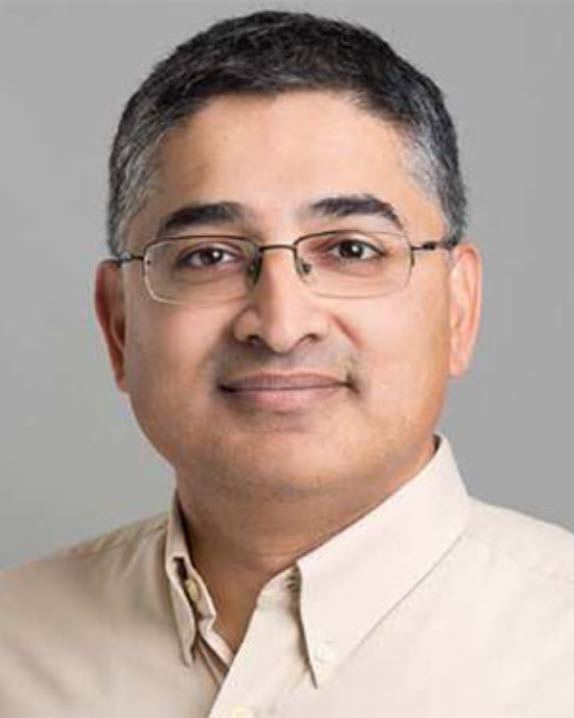}}]
 {Raviraj Adve} (Fellow, IEEE) was born in Bombay, India. He received the B.Tech. degree in electrical engineering from IIT, Bombay, in 1990, and the Ph.D. degree from Syracuse University, NY, USA, in 1996, His thesis received the Syracuse University Outstanding Dissertation Award. From 1997 to August 2000, he worked for Research Associates for Defense Conversion Inc. on contract with the Air Force Research Laboratory at Rome, NY. He joined the Faculty with the University of Toronto in August 2000 where he is currently a Professor. Dr. Adve’s research interests include analysis and design techniques for cooperative and heterogeneous networks, energy harvesting networks and in signal processing techniques for radar and sonar systems. He was the recipient of the 2009 Fred Nathanson Young Radar Engineer of the Year award.
 \end{IEEEbiography}
  \vspace{1pt}
 \begin{IEEEbiography}[{\includegraphics[width=1in,height=1.25in,clip,keepaspectratio]{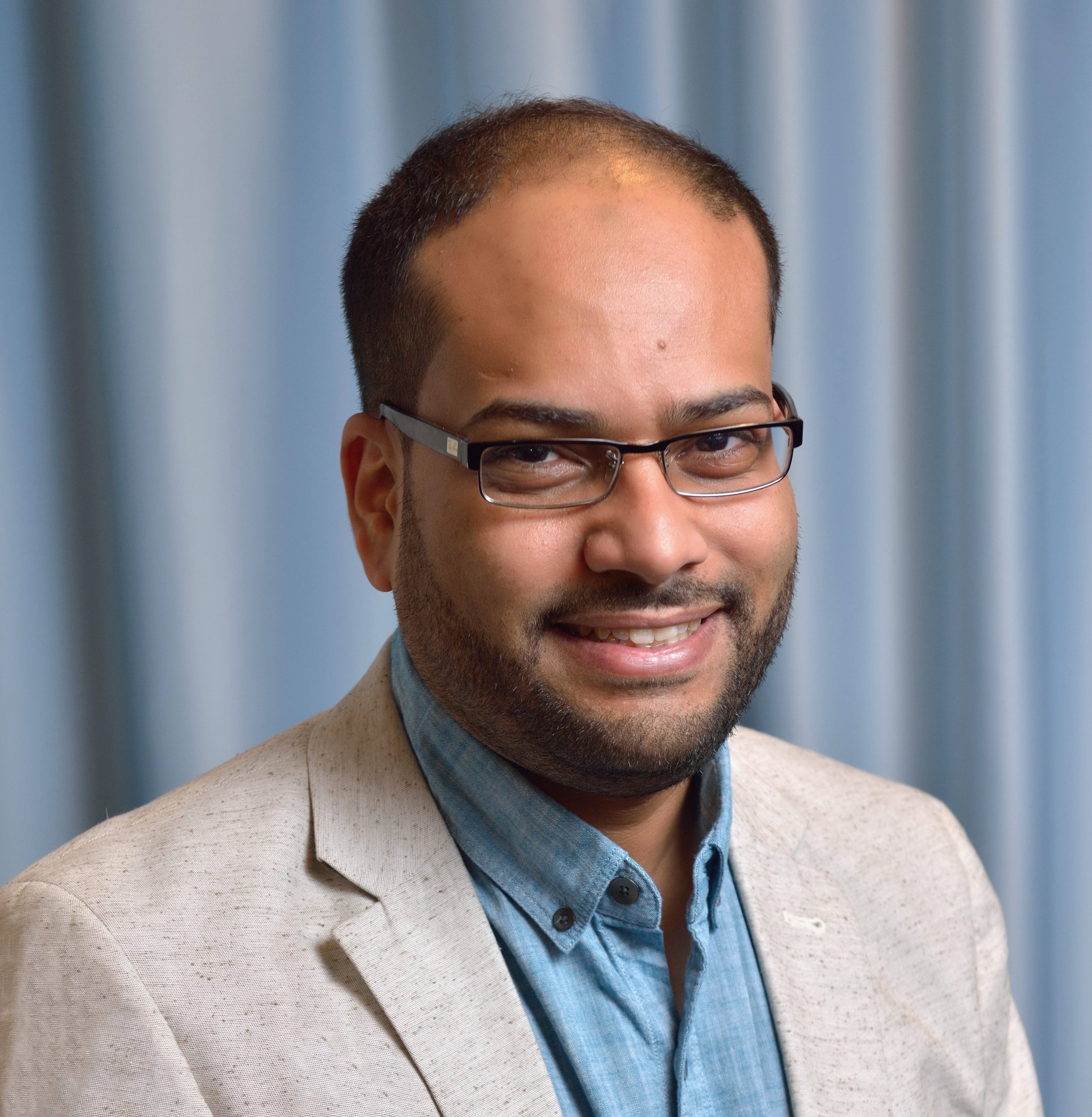}}]{Dr. Akram Bin Sediq}
	 is a Principal AI/ML Developer at Ericsson Canada, where he leads the AI/ML technical development in physical-layer and radio-resource-management domains. His work has resulted in 40+ granted and filed patents, as well as 50+ peer-reviewed publications. Dr. Bin Sediq is a recipient of the President's Cup for graduating with the highest GPA at the Bachelor’s level at AUS in 2006, the Ontario Graduate Scholarship for international students for three years in a row during 2007–2010, two Senate Medals for outstanding academic achievement from Carleton University at the Master's and Ph.D. levels in 2008 and 2013, respectively, the EDC Teaching Assistant Outstanding Award at Carleton University in 2013, NSERC Industrial R\&D Fellowship (IRDF) from 2013 to 2015, Ericsson's Key Contributor Award for three years in a row (2015-2017), AUS Alumni Wall of Fame recognition in 2015, and a Best Paper Award at IEEE ICC 2022.
\end{IEEEbiography}
 \vspace{1pt}
\begin{IEEEbiography}[{\includegraphics[width=1in,height=1.25in,clip,keepaspectratio]{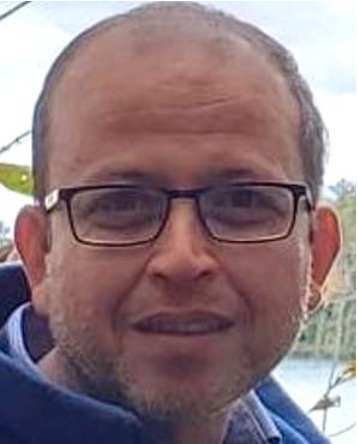}}]{Amr El-Keyi}
	(Member, IEEE) received the B.Sc. and M.Sc. degrees in electrical engineering from Alexandria University, Alexandria, Egypt, in 1999 and 2002, respectively, and the Ph.D. degree in electrical engineering from McMaster University, Hamilton, ON, Canada, in 2006. From 2009 to 2015, he was an Assistant/Associate Professor with the School of Communication and Information Technology, Nile University, Giza, Egypt. He is
	currently a System Developer at Ericsson Canada	Inc. His research interests include array signal processing and interference management for wireless communication systems.
\end{IEEEbiography}
\end{document}